\documentstyle[12pt]{article}
\begin{document}

\centerline{\bf\Large Escape of VHE gamma-rays from close massive binary Cen
X-3}
\vskip 0.5truecm
\centerline{\large W. Bednarek}
\vskip 0.5truecm
\centerline{\it\large Department of Experimental Physics, University of 
\L \'od\'z,}
\vskip 0.35truecm
\centerline{\it\large ul. Pomorska 149/153, PL 90-236 \L \'od\'z, Poland }

\begin{abstract}
{We consider propagation of very high energy (VHE) gamma-rays in the
radiation field of a massive star in the  binary system Cen X-3, 
which has been reported as a source  of gamma-ray photons in the GeV and TeV 
energies. VHE gamma-rays or electrons, injected by the compact object, 
should develop inverse 
Compton $e^\pm$ pair cascades. Based on the Monte Carlo simulations we
predict the $\gamma$-ray spectra and the $\gamma$-ray light curves for 
the parameters of Cen X-3 system, and applying different
models of  injection of primary particles (monodirectional, isotropic,
monoenergetic,  power law). It is found that the gamma-ray spectra 
observed at different directions  have different shape  and intensity.
Interestingly, some  gamma-rays may even escape
from the binary system at directions which are obscured by the massive 
star, i.e. from the opposite side of the massive star than 
the location of the compact source of primary gamma-rays or electrons. 
We predict that the gamma-ray light curves, produced 
in the case of electron injection by the compact object in the Cen X-3
system, should have opposite tendencies for photons with energies above 
100 MeV and above 300 GeV, i.e. the photon intensities increases with
phase in the first case and decreases with phase in the second case. 
However model with injection of primary electrons seems to be in 
contrary with the reported modulation  of the GeV $\gamma$-ray flux  
with the pulsar's period. 
The model with injection of primary photons allows such modulation
with the pulsar's  period, but predicts strong modulation of the TeV 
flux
with the  orbital period of the binary. Modulation of TeV emission 
with the orbital period has been 
reported by the early Cherenkov observations, but was not confirmed
by the recent, more sensitive observations by the Durham Mark 6 
telescope. } 

\end{abstract}

\section{Introduction}

The massive binary systems are often considered as sites of high energy
processes in which $\gamma$-ray production is expected. $\gamma$-rays in
these systems may be  produced in
interactions of relativistic particles injected by: a neutron star or 
a black hole (e.g. Bignami, Maraschi \& Treves~1977, Kirk, Ball \& 
Skjaeraasen~1999);
a shock wave  created in collision of the pulsar and stellar winds (e.g.
Harding \& Gaisser~1990) or two stellar winds (e.g. Eichler  \& 
Usov~1993). Observations of these systems in GeV and TeV energies 
suggest that in fact it may be the case. For example in last years the
Compton GRO detectors reported point sources with flat spectra 
coincident with some massive binaries e.g. Cyg X-3, 
LSI 303$^{\rm o}$+61, or Cen X-3. Some positive detections of massive
binaries by Cherenkov telescopes has been also claimed, although they 
were frequently accompanied by many negative reports (see for a 
review Weekes~1992, Moskalenko~1995). In the recent review on TeV 
observations only Cen X-3 has been mentioned as possible massive 
binary active at these energies (Weekes~1999).

These theoretical predictions and positive observations has stimulated
analysis of propagation of VHE $\gamma$-rays in the anisotropic 
radiation fields of accretion disks around compact objects in massive
binaries (Carraminana~1992, Bednarek~1993), and in the radiation 
fields of massive stars (e.g. Protheroe \& Stanev~1987, Moskalenko, 
Karaku\l a \& Tkaczyk~1993, Moskalenko \& Karaku\l a~1994).  
Recently we have performed Monte Carlo simulations of cascades initiated
by monoenergetic $\gamma$-ray beams injected by a discrete source
(probably a compact object) in the radiation field of a
massive companion. Two massive binaries,  
Cyg X-3 (Lamb et al.~1977, Merck
et al.~1995) and  LSI 303$^o$+61 (van Dijk et al.~1993, Hermsen et al.~
1977, Thompson
et al.~1995), have been discussed in the context of this problem
(Bednarek~1997).

Another massive X-ray binary, Cen X-3, containing a neutron
star with the period 4.8 s on a 2.09 day orbit around an O-type 
supergiant, has been detected above $100$ MeV by the EGRET detector on 
the Compton Observatory (Vestrand, Sreekumar \& Mori~1997).
There are evidences that $\gamma$-ray emission at these 
energies has a form of outbursts and is modulated with a 4.8 s period 
of the pulsar. Based on  the observations in late 80's,
two Cherenkov groups (Raubenheimer et al.~1989, 
Brazier et al.~1990, North et
al.~1990) reported detection of positive signal at TeV energies from Cen X-3
at an orbital  phase of $\sim 0.75$, which is modulated with a period of the
pulsar.  This emission has been localized by North et al.~(1991) and 
Raubenheimer 
\& Smith~(1997) to a relatively small region between the pulsar orbit 
and the surface of a massive companion which may be the accretion 
wake or the limb of the star. More recently the Durham group has 
detected
a persistent flux of $\gamma$-rays above 400 GeV on a lower level than
previous reports (Chadwick et al.~1998,1999). No evidence of correlation 
with the pulsar or orbital periods has been found and no evidence of 
correlation with the X-ray flux has been detected (Chadwick et 
al.~1999a,b).

The purpose of this paper is to compute the $\gamma$-ray light curves
and $\gamma$-ray spectra escaping from Cen X-3 system, assuming 
different geometries and energy distributions for primary
electrons or photons injected into the radiation field of the 
massive star by a compact object. 
%The geometry of the considered picture is schematically shown in 
%Fig1.
The $\gamma$-ray spectrum escaping from the system is a consequence of 
anisotropic cascades which initiate primary electrons or 
$\gamma$-ray photons in the thermal radiation of a massive star in 
Cen X-3. The results  of computations are discussed in the context of 
observations of Cen X-3 in GeV and TeV energy ranges.  

We assumed the following parameters for the massive binary Cen X-3: the 
radius of the O star $r_{\rm s} = 8.6\times 10^{11}$ cm, its surface
temperature $T_{\rm s} = 3\times 10^4$ K (Krzemi\'nski~1974), the binary star
separation $a = 1.4 r_{\rm s} = 1.2\times 10^{12}$ cm, the inclination 
angle
$70^o$ and  the circular orbit since the known eccentricity of the binary is 
$< 0.0008$ (Fabbiano \& Schreier~1977).

%%___________________________ 
%\begin{figure} 
%  \vspace{4.5cm} 
%\special{psfile=cenfign0.eps voffset=-20 hoffset=0 hscale=40 vscale=35}
%  \caption[]{Schematic picture of cascade initiated by primary gamma-ray
%($\gamma_1$) in the radiation field of a massive star. Gamma-ray,
%injected by the compact object or produced by primary electron, 
%creates an $e^\pm$ pair in the interaction with a soft star photon
%$\varepsilon$. The secondary $e^\pm$ pairs create secondary gamma-rays 
%by scattering soft star photons (one of them is marked by $\gamma_2$). 
%The next generation of gamma-rays may create: next $e^\pm$ pairs; escape 
%from the system (e.g. photon marked by $\gamma_4$); or collide with the 
%star surface.}      
%\end{figure} 
%%___________________________ 

%
%
\section{Propagation of VHE $\gamma$-rays in the radiation field
of a massive star}

In the previous paper (Bednarek~1997), we considered the propagation of
monoenergetic $\gamma$-ray beams in the radiation filed of a massive
star in the case of two massive binaries: Cyg X-3 and LSI 303$^o$+61. 
We determined general conditions for which VHE $\gamma$-rays may 
initiate such kind of inverse Compton cascade (ICS) (see Sect. 
2 in Bednarek~1997). The massive star in Cen X-3 system has also high
enough luminosity that VHE photons injected inside a few stellar radii
should initiate the cascade. This expectation has been checked by 
computing the optical depth for $\gamma$-ray photons as a function of 
distance from the surface of the massive star, the photon energies, 
and the angle of injection $\alpha$ measured  with 
respect to the direction determined 
by the place of  photon injection  and the center of the massive star. 
The optical depth  has been computed as shown in Appendix A of
Bednarek~(1997). We have found  that for the parameters of the massive 
star in Cen X-3 system,  the optical depth for photons injected at 
angles $\alpha > 30^o$ and at the  distance $x = 1.4r_{\rm s}$, can be 
greater than one.  
If photons are injected closer to the massive star, then the optical 
depth can be greater than one for all injection angles 
(see curve for $x = 1.1r_{\rm
s}$ in Fig. 1a). The optical depth reaches maximum 
(characteristic cusps visible in Fig. 1a),  for photons
propagating in directions tangent to the massive star surface, i.e.
at an angle $\alpha_\delta = \pi - \delta$, where $\delta$ is the 
angle intercepted by the massive star. For higher angles $\alpha$, the 
optical depth is computed up to the moment of photon collision with the 
stellar surface and therefore it is lower than for $\alpha_\delta$.
As expected the optical depth for $\gamma$-ray photon
reaches maximum at photon energies 
determined by characteristic energies of soft star photons (see 
Fig. 1b).
Photons with energies below a few tens of GeV escape without significant
absorption. Therefore it is expected that cascade $\gamma$-ray spectra 
should show a break close to these energies.

%___________________________ 
\begin{figure} 
  \vspace{10cm} 
\includegraphics{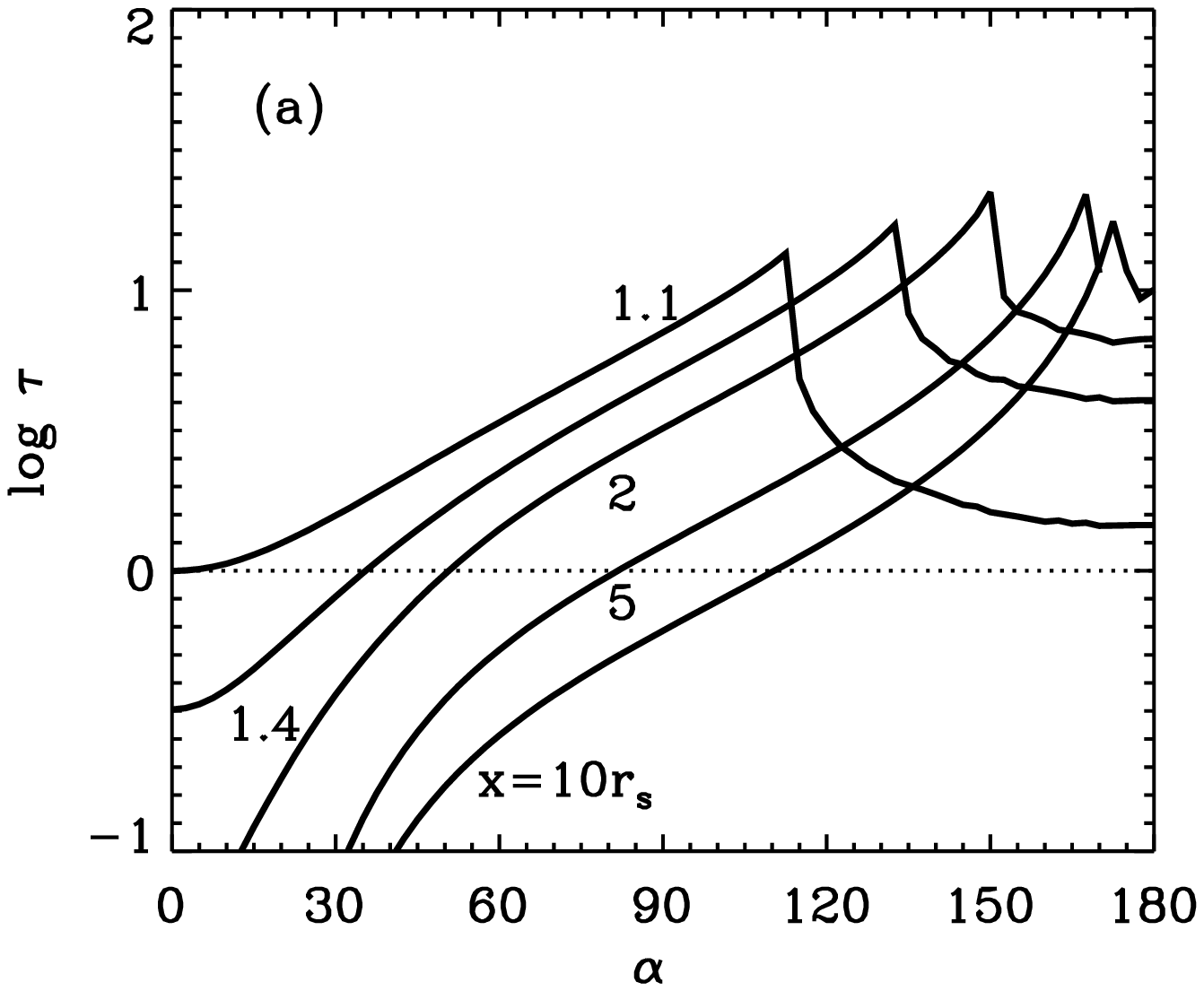}
\includegraphics{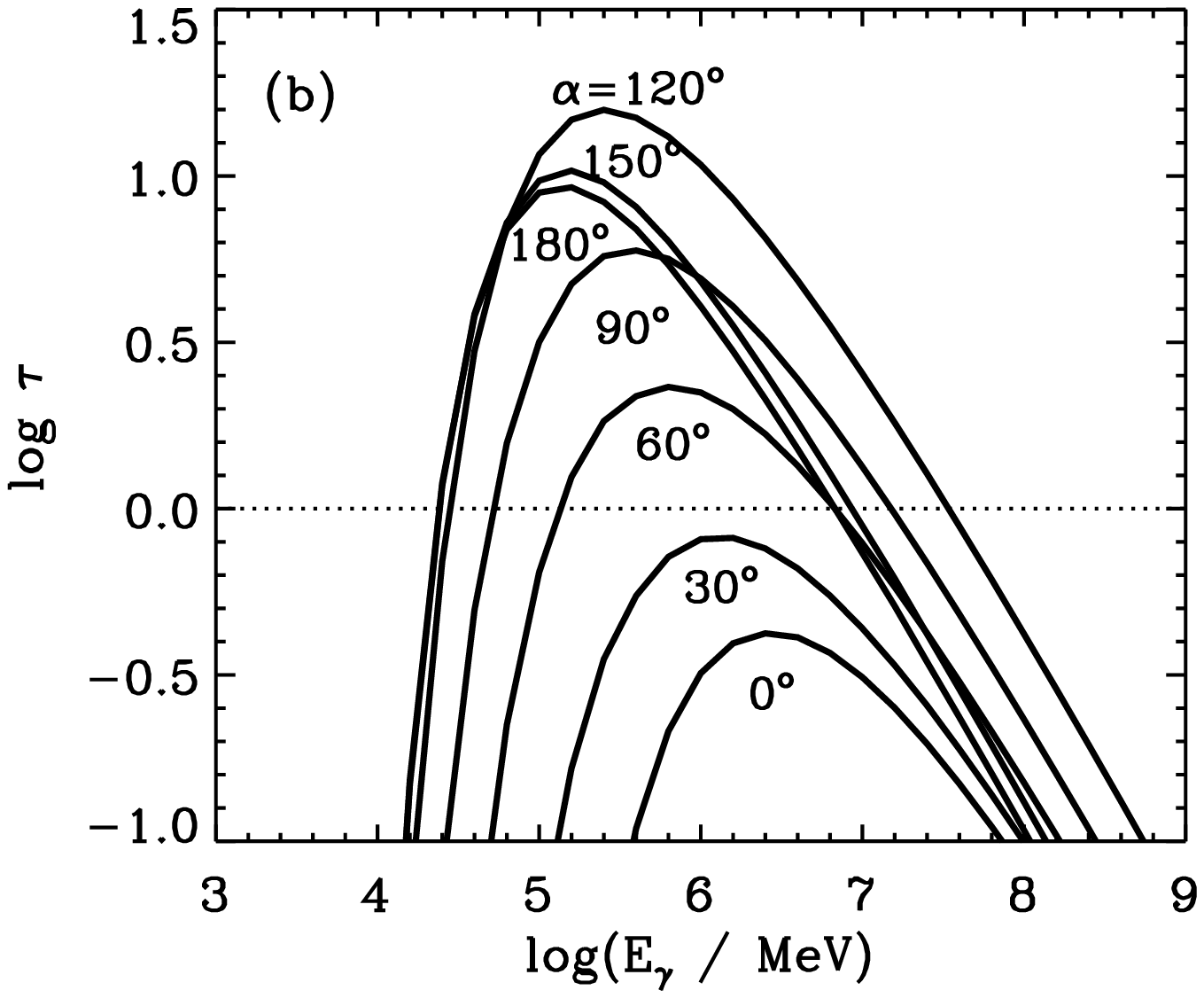}
  \caption[]{The optical depth for $\gamma$-ray photons with 
energies $10^{6}$ MeV in the soft radiation field of the massive star in
Cen X-3 system is shown as a function of the angle of injection of photons
$\alpha$, for different distances of the photon injection
from the massive star equal to $x = 1.1, 1.4, 2,5, 10$ radii of the star
$r_{\rm s}$ {\bf (a)}. The dependence of the optical depth on the photon 
energy, for fixed values of the angle $\alpha$ and the distance $x = 
1.4r_{\rm s}$, is shown in {\bf (b)}.}    
\end{figure} 
%___________________________ 

% 
%
\section{Gamma-ray spectra from ICS cascade} 

Let us assume that the $\gamma$-ray photon is injected by the compact
object which is on an orbit around the star with the parameters 
characteristic for Cen X-3 system. As we showed above these photons may 
create $e^\pm$ pair in collision with the soft star photon.
The secondary pairs can next produce ICS $\gamma$-rays, initiating in 
this way the ICS cascade in the massive star radiation which 
is seen anisotropic in
respect to the location of the injection place of the primary photons
or electrons. In the subsections we describe the cascade scenario and
discuss the results of calculations for different initial injection
conditions, i.e. type, distribution, and spectra of primary particles.

\subsection{ICS cascade with isotropisation of secondary electrons}

The $\gamma$-ray photon with energy $E_\gamma$ interacts with the 
star photon creating $e^{\pm}$ pair at the propagation distance 
$l_\gamma$. The place of the creation can be simulated randomly from 
\begin{eqnarray}
P_1 & = & {\rm exp}\left({-\int_0^{\rm l_\gamma}
\lambda^{-1}_{\gamma\gamma}(E_\gamma,x_\gamma,\alpha_\gamma)}dl\right)
\end{eqnarray}
\noindent
where $l$ is the distance measured along the photon propagation, 
$\lambda_{\gamma\gamma}(E_\gamma,x_\gamma,\alpha_\gamma)$ is the mean 
free path for $\gamma$-ray photon with energy $E_\gamma$, injected at 
the angle $\alpha_\gamma$ and at the distance $x_\gamma$ from the 
center of the massive star in Cen X-3 system, and $P_1$ is the random 
number. If condition~(1) can not  be fulfilled for chosen random number 
$P_1$, we accept that the $\gamma$-ray: escapes from the star radiation 
field, or collides with the star surface.

%
%___________________________ 
\begin{figure*} 
  \vspace{10cm} 
\includegraphics{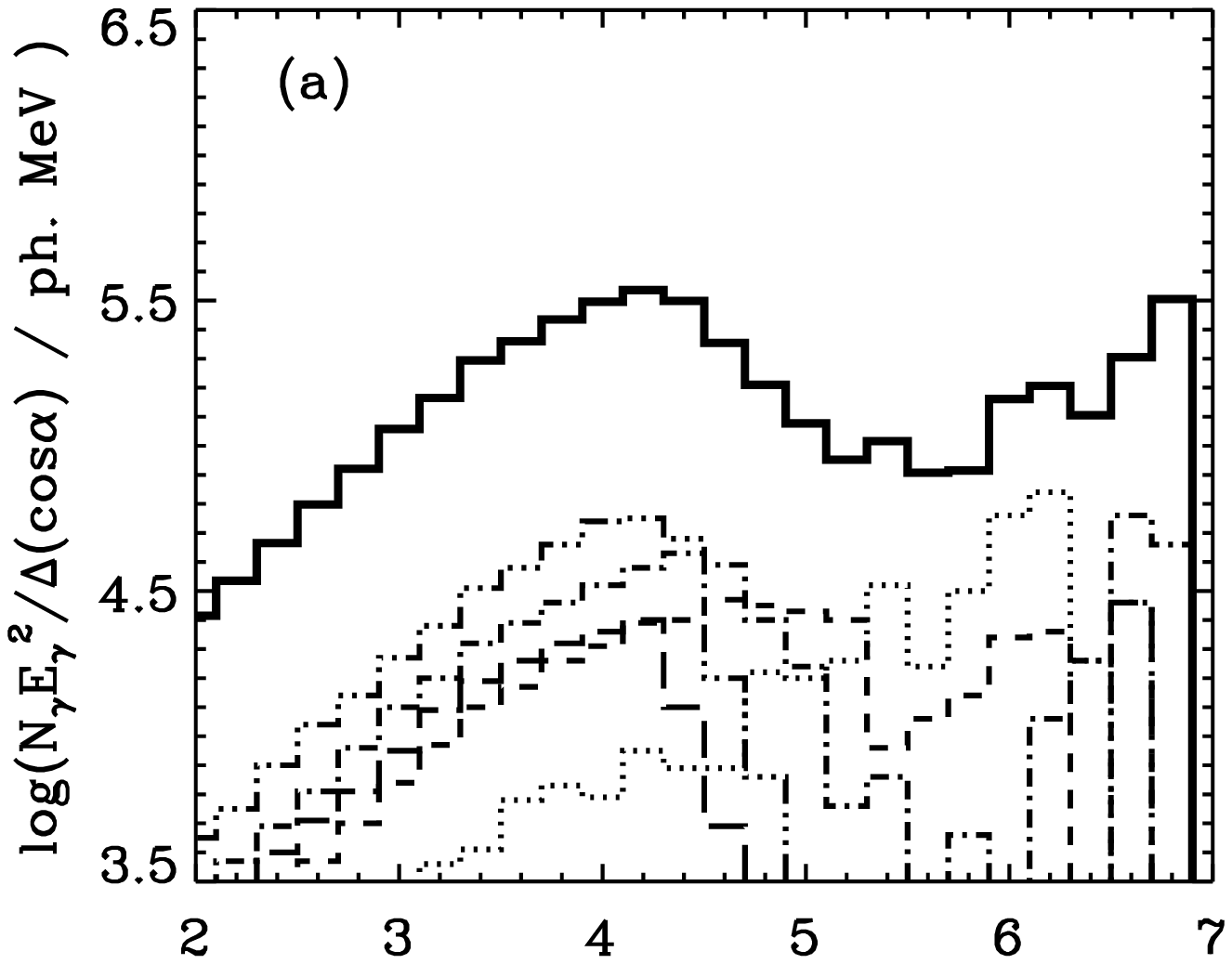}
\includegraphics{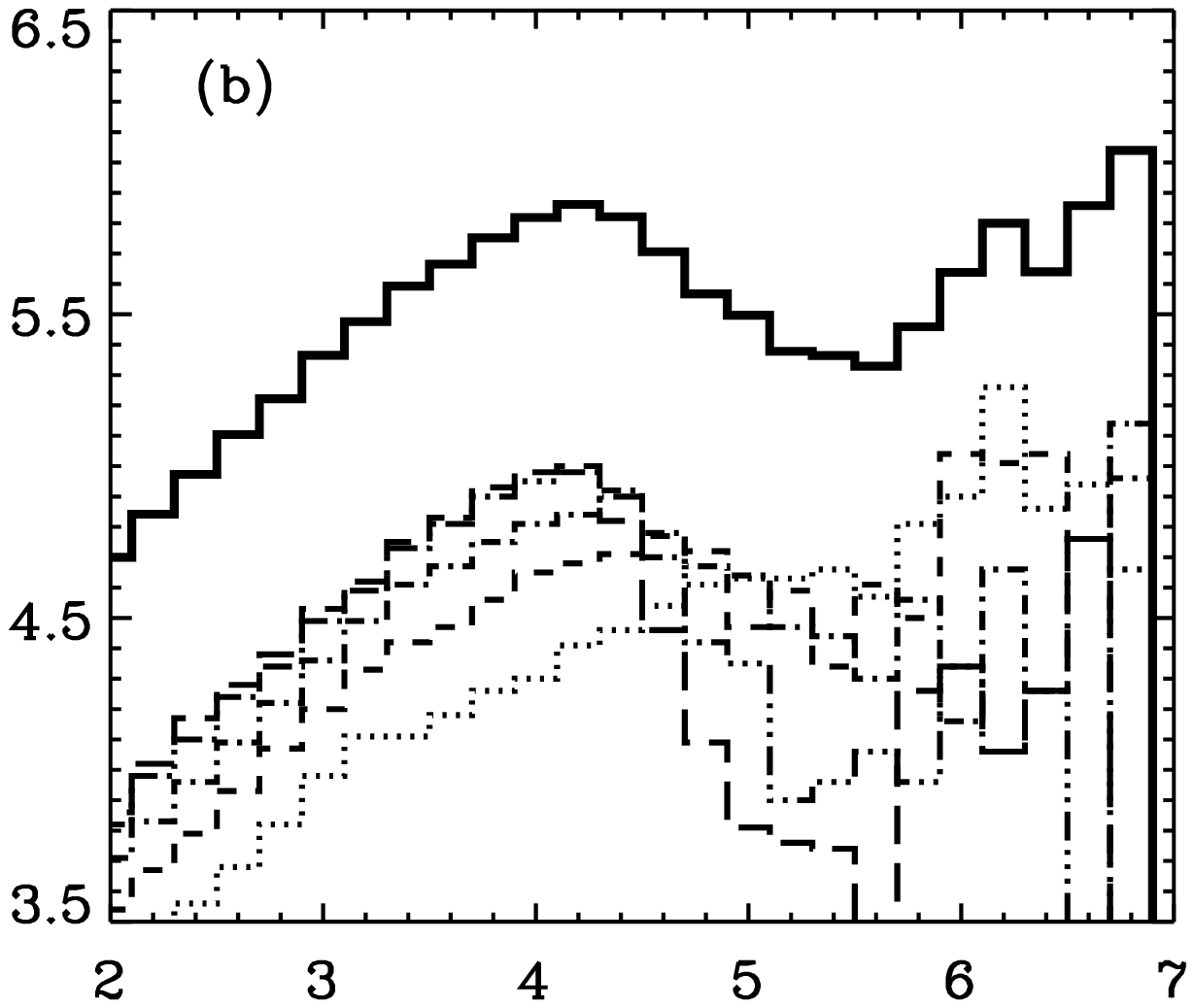}
\includegraphics{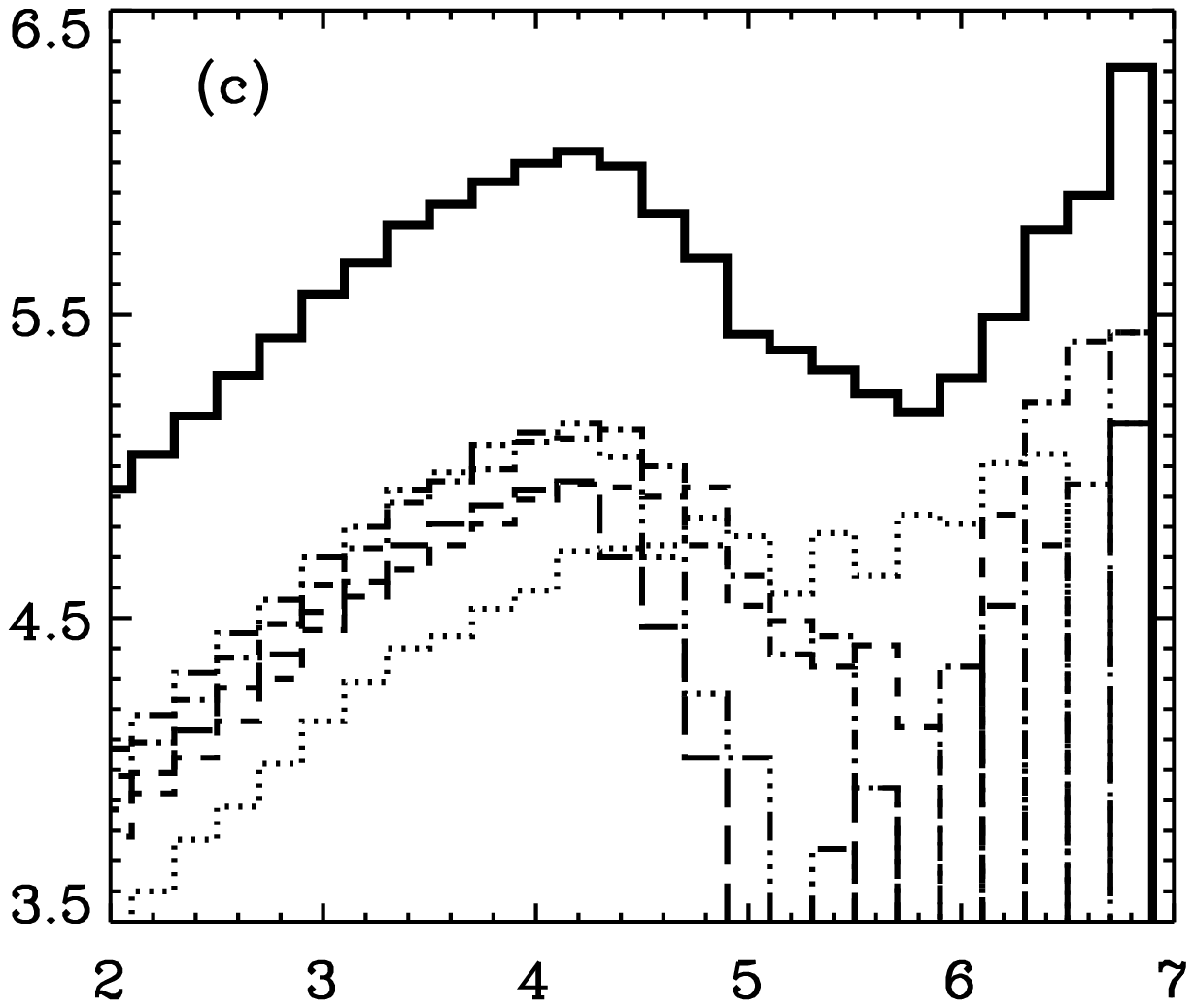}
\includegraphics{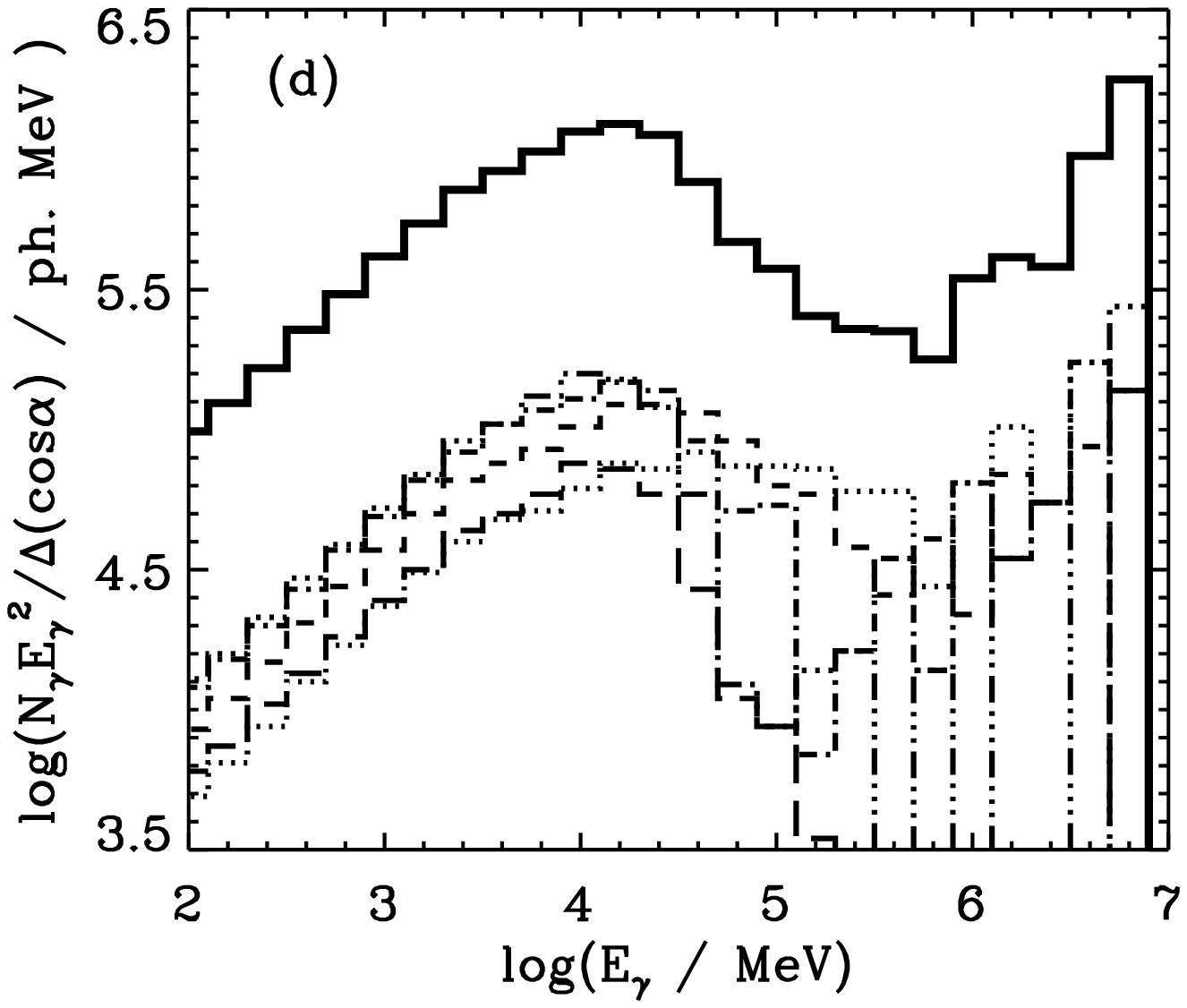}
\includegraphics{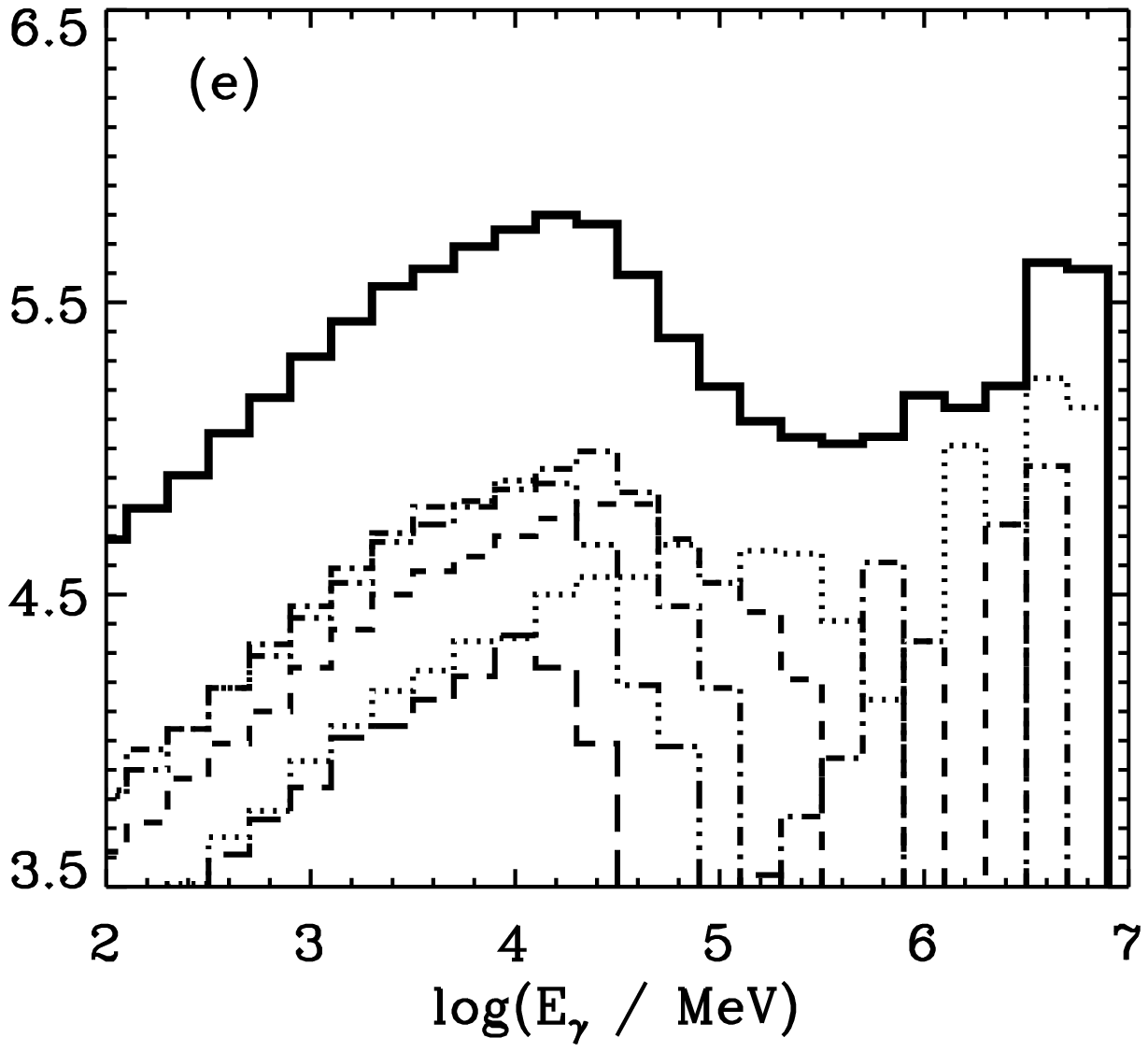}
\includegraphics{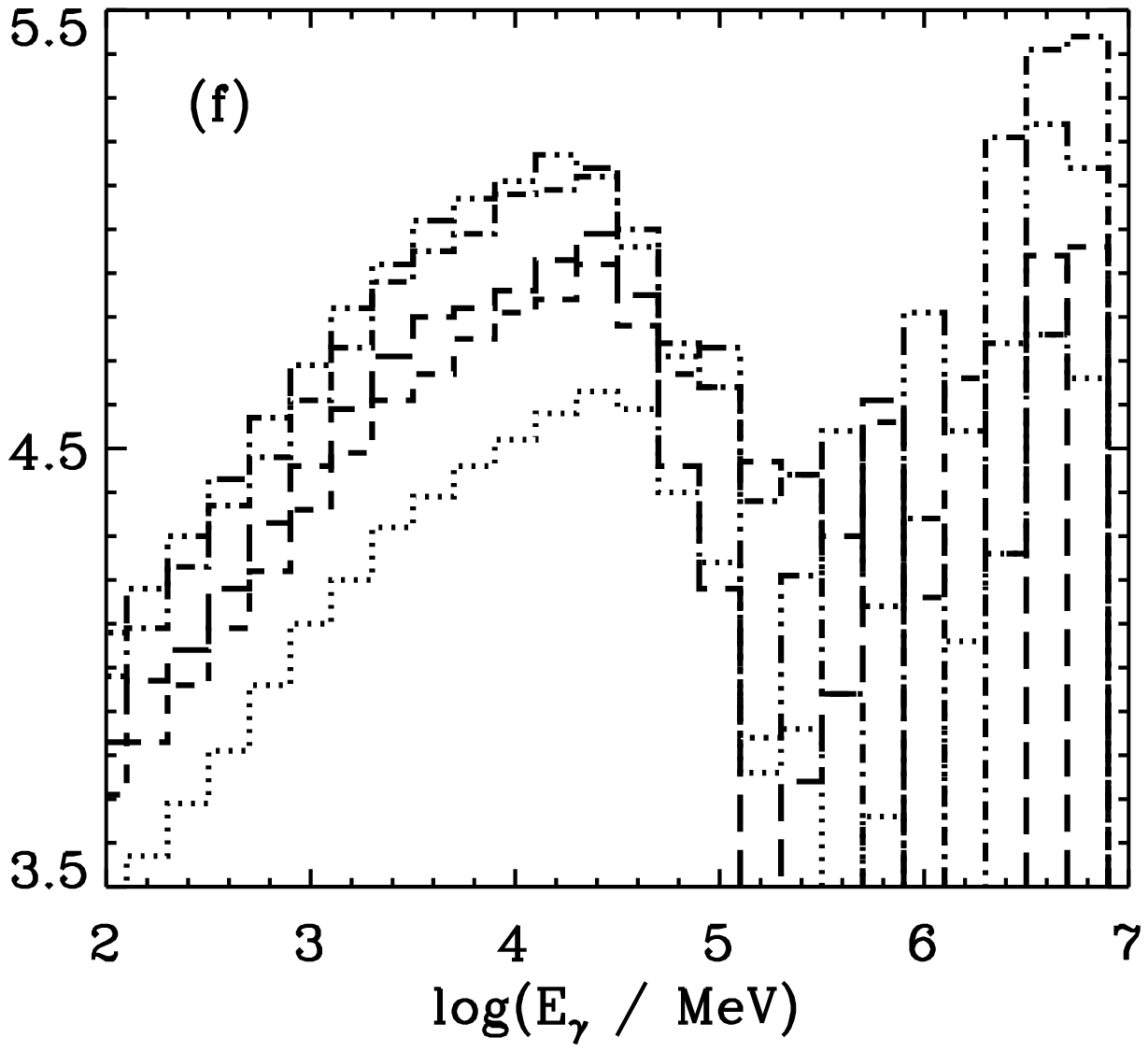}
\caption[]{The secondary spectra of escaping $\gamma$-rays from
the binary  system
with Cen X-3 parameters, in the case of injection of monodirectional and
monoenergetic primary photons with energy $10^7$ MeV. The spectra are shown
for  different 
angles of injection of primary $\gamma$-rays: $\cos\alpha = 1$ {\bf (a)}, 
0.5 {\bf (b)}, 0. {\bf (c)}, $-0.5$ {\bf (d)}, and $-1$ {\bf (e)}, within the
range of  observation angles: 
$\cos\theta = 0.8 \div 1.$ (dotted histograms), $0.4 \div 0.6$ (dashed),
$0.2 \div 0$ (dot-dashed), $-0.6 \div -0.4$ (dot-dot-dashed), and
$-1 \div -0.9$ (long-dashed). The thick solid histogram shows the
secondary $\gamma$-ray spectrum integrated over all sphere.  
In figure {\bf (f)} we compare the spectra escaping at fixed range of angles 
$\cos\theta = 0 \div 0.2$ in the case of different angles of 
injection of  primary photons $\cos\alpha = 1.$ (dotted 
histogram), 0.5 (dashed),  0. (dot-dashed), $-0.5$ (dot-dot-dot-dashed), 
and $-1.$ (long dashed). } 
\end{figure*} 
%___________________________ 
%

Simple trigonometry allows us to determine the distance of produced 
$e^\pm$ pair, $x_{e^{\pm}}$, from the center of the massive star. 
The energy of created electron (positron) $E_{\rm e}$ is chosen by 
sampling
from the differential spectrum of pairs which can be produced by the
$\gamma$-ray photon at the propagation distance $l_\gamma$.
It is obtained from
\begin{eqnarray}
P_2 & = & \left(\int_{0.5}^{E_e} {{dW}\over{dEdx}}dE\right)
\left( \int_{0.5}^{E_{e,max}}{{dW}\over{dEdx}}dE\right)^{-1}
\end{eqnarray}
\noindent
where $dW/dEdx$ is defined in Bednarek~(1997, Appendix B),  
$E_{e,max}$ is the  maximal possible energy of the electron produced in
$\gamma$-ray -- soft photon  collision, and $P_2$ is the random number. 
The energy of positron is then $E_p = E_\gamma-E_e$.

We assume that secondary $e^{\pm}$ pairs 
are locally isotropised by the random component of the magnetic field.
This will happen if the random component of the magnetic field 
$B_{\rm r}$ inside the binary system is strong enough (Bednarek~1997), 
i.e.
\begin{eqnarray} 
B_r \ge 4\times 10^{-29} \gamma_{\rm e}^2 T^4D,
\end{eqnarray} 
where $\gamma_{\rm e}$ is the Lorentz factor of electron 
(or positron), 
$T$ is the surface temperature of the massive star, and $D$ is the 
dilution factor of the star radiation defined as the part of the sphere 
intersepted by the star. $D\approx 0.1$ for the separation of the 
companions $x = 1.4 r_{\rm s}$.  
The $e^\pm$ pairs cool locally and produce next generation of
$\gamma$-rays in the inverse Compton scattering of soft photons coming 
from the massive  star. In order to select the direction of the
secondary $\gamma$-ray we compute at first the energy loss rate on ICS 
for pairs with energy $E_{e^{\pm}}$ assuming that the electron (positron) 
is located at the distance $x_{e^{\pm}}$ and propagates at an angle
$\alpha_{e^\pm}$ to the direction defined by $x_{e^{\pm}}$ (see 
Appendix C in Bednarek~1997). 
The direction of motion of a secondary $\gamma$-ray, which in fact
covers with the direction of motion of relativistic electron at the 
moment of $\gamma$-ray production, is obtained by sampling from the 
distribution of the energy loss rates  computed as a function of the 
angle $\alpha_{e^\pm}$ after its normalization to the total energy 
loss rate for electrons moving isotropically at $x_{e^{\pm}}$. 
The cosine of the angle of $\gamma$-ray emission ($\cos\alpha_\gamma$) 
is obtained from
\begin{eqnarray}
P_3 & = & 
\left(\int_{\cos\alpha_\gamma}^1 {{dL}\over{dtd\nu}}d\nu\right)
\left(\int_{-1}^1 {{dL}\over{dtd\nu}}d\nu\right)^{-1},
\end{eqnarray}
\noindent
where $P_3$ is the random number, and
\begin{eqnarray}
{{dL}\over{dt d\nu}} & = & \int_0^{E_{\gamma, max}} 
{{dN}\over{dt dE_\gamma}}
{{dN_{e^{\pm}}}\over{d\nu}}E_\gamma dE_\gamma, 
\end{eqnarray}
\noindent
$dN/dt dE_\gamma$ is the photon spectrum produced by electrons in 
ICS process (see Appendix C in Bednarek~1997), 
$dN_{e^\pm}/d\nu$ describes the number of electrons moving inside the
unit cosine angle $d\nu = d(\cos\alpha)$, and $E_{\gamma,max}$ is the 
maximum energy of the $\gamma$-ray photon
produced in the ICS process by an electron with energy $E_{e^\pm}$, 
located at the
distance $x_{e^\pm}$, and propagating at the angle 
$\alpha_{e^\pm} =\alpha_\gamma$.
%The azimuthal angle $\phi$ of the photon in spherical coordinates is 
%given by $\phi  =  2\pi P_4$, where $P_4$ is the next random number.
 In this way the directions of 
motion of secondary ICS $\gamma$-rays are determined. Now we need to 
determine their energies. The mean
energy of secondary $\gamma$-ray photon, $E'_\gamma$, produced by 
secondary electron
with energy $E_{e^{\pm}}$ at the distance $x_{e^{\pm}}$ from the star 
and at an angle $\alpha_\gamma$, is obtained from the formula
\begin{eqnarray}
<E_\gamma> & = \left({{dL}\over{dt d\nu}}\right)
\left({{dN}\over{dt d\nu}}\right)^{-1},
\end{eqnarray}
\noindent
where
\begin{eqnarray}
{{dN}\over{dt d\nu}} & = & \int_0^{E_{\gamma, max}} 
{{dN}\over{dt dE_\gamma}}
{{dN_{e^{\pm}}}\over{d\nu}} dE_\gamma 
\end{eqnarray}
\noindent
is the rate of photon production by an electron in the ICS process.
%The dependence of $<E_\gamma>$ on $E_{e^{\pm}}$ is shown in Figs.~4ab
%in Bednarek~(1997).
Above described procedure is repeated for all cascade $e^\pm$ pairs up 
to the 
moment of their 'complete cooling', i.e. to the moment at which they are
not able to produce $\gamma$-ray photons in ICS process with energies
above certain applied value which in our simulations is chosen as equal 
to 100 MeV. We follow all secondary ICS
$\gamma$-rays with energies above the threshold  for $e^\pm$ pair 
production in radiation of a massive star. 

\subsection{Injection of monodirectional and monoenergetic photons
or electrons}

%___________________________
\begin{figure} 
   \vspace{6.cm} 
\includegraphics{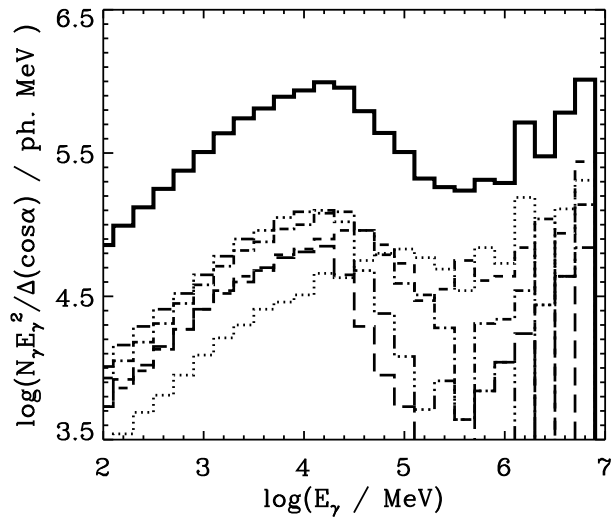}
\caption[]{The gamma-ray spectra escaping from the Cen X-3 binary 
system after cascading in the soft radiation field of the massive star 
in the case of isotropic injection of monoenergetic $\gamma$-ray 
photons with energy $E_\gamma = 10^7$ MeV. The secondary photon spectra  
are observed at a range angles: $\cos\theta = 0.8 \div 1.$ 
(dotted),  $0.4 \div 0.6$ (dashed), $0 \div -0.2$ (dot-dashed), 
$-0.6 \div -0.4$
(dot-dot-dot-dashed),  $-1 \div -0.8$ (long-dashed). The full thick 
histogram shows the secondary spectrum summed over all sphere.}
\end{figure} 
%___________________________ 

It is likely that high energy photons and/or electrons are injected 
by the neutron star (a compact object in Cen X-3 system) in the form 
of highly collimated beams. Such beams may be formed: in the outer 
gaps of pulsar magnetospheres (e.g. Cheng, Ho \& Rudermann~1986); in
collisions of  collimated
protons escaping from the pulsar magnetospheres with the matter of an 
accretion disk (Cheng \& Ruderman~1989); or they may emerge from 
regions of the magnetic poles of the neutron star (Kiraly \& 
Meszaros~1988). In our simulations it is assumed that highly collimated 
beam of $\gamma$-rays with energy 10 TeV emerge from the distance
$x = 1.4 r_{\rm s}$ at a certain
angle $\alpha$, measured with respect to the direction determined by the
center of the massive  star and the compact object. In fact
such situation corresponds also to  the case of the highly collimated 
electron beams with comparable energy since at these 
energies electrons transfer almost all its primary energy to a single 
$\gamma$-ray photon in ICS process because the scattering occurs in the
Klein-Nishina domain.

We are interested in the spectra of secondary photons which escape from 
the above described type of cascade at a range of angles 
$\Delta\cos\theta = 
0.2$. In Fig. 2, we show such secondary spectra 
for different injection angles of primary monoenergetic photons with 
energy $E_\gamma = 10^7$ MeV: $\cos\alpha = 1$ (a), 0.5 (b), 0. (c),
$-0.5$ (d), and $-1$ (e), observed at the range of angles: 
$\cos\theta = 0.8 \div 1.$ (dotted histograms), $0.4 \div 0.6$ (dashed),
$0.2 \div 0$ (dot-dashed), $-0.6 \div -0.4$ (dot-dot-dashed), and
$-1 \div -0.9$ (long-dashed). The cascading effects are the strongest 
(the highest numbers of secondary $\gamma$-rays) if the monoenergetic
$\gamma$-ray beam is injected at the intermediate angles $\alpha$
(see thick solid histograms in Fig. 2, which show the 
secondary photon spectra intergated over all solid angle). This is a
consequence of the highest optical depth for photons propagating at 
directions tangent to the massive star limb (curve for $x = 1.4r_{\rm s}$ 
in Fig. 1a). All secondary spectra are well described by a 
power law with the index  $\sim 1.5$ at energies below $\sim 10$ GeV 
(as expected in ICS cascades with complete cooling of  electrons). The 
photon intensities in these spectra are the highest at direction
tangent to the limb of the massive star (dot-dot-dot-dashed histograms
in Fig. 2). At
higher photon energies ($> 10$ GeV), the spectra show characteristic 
cut-offs due to the absorption of secondary photons in the massive star
radiation. The spectra recover at TeV energies with the 
intensities depending on the observation angle. The highest intensities 
are observed at small angles $\alpha$ for which the optical depth for 
TeV $\gamma$-rays is the lowest (see Fig. 1).
In Figure 2f, we compare the spectra escaping at  
fixed range of angles $\cos\theta = 0 - 0.2$ but for different angles 
of injection of  primary photons $\cos\alpha = 1.$ (dotted 
histogram), 0.5 (dashed),  0. (dot-dashed), $-0.5$ (dot-dot-dot-dashed), 
and $-1.$ (long dashed). It is clear that primary photons injected
at directions of the highest optical depth ($\cos\alpha = -0.5 \div 0.$)
produce the secondary photons with the highest intensities than
thes onces injected at directions of the lowest optical depth
($\cos\alpha = 1.$).

Note that in the type of cascade discussed here, the intensity of
secondary photons observed at the direction of injection of primary 
photons
is not completely dominated by these primary photons because of the
large optical depths for the considered injection place ($x = 1.4 
r_{\rm s}$), and the isotropisation of secondary cascade $e^\pm$ pairs.
Our assumptions on the propagation of photons in massive binaries are
different than these ones applied in e.g. Kirk, Ball \& 
Skjaeraasen~(1999). 

\subsection{Isotropic injection of monoenergetic photons or electrons}

The monoenergetic photons and electrons can be isotropically injected
into the binary system by the young pulsars with strong pulsar winds. 
In fact, it is believed  the pulsar winds are composed with relativistic
electrons (positrons) which have typical Lorentz factors of the order 
$\sim 10^7$ (Rees \&  Gunn~1974). These electrons, if accelerated by the 
electric fields generated during magnetic reconnection in the pulsar 
wind zone, can be injected approximately isotropically. Therefore 
we consider the case of isotropic injection of photons or electrons
with energies 10 TeV. If only 10 TeV electrons are injected, they should 
produce photons with comparable  energies in a single scattering as 
mentioned above. The secondary photon spectra, produced in cascade under 
such initial conditions, are shown in Fig. 3 for different 
range of observation angles defined by: $\cos\theta = 0.8 \div 1.$ 
(dotted),  $0.4 \div 0.6$ (dashed), $0 \div 0.2$ (dot-dashed), 
$-0.6 \div -0.4$
(dot-dot-dot-dashed),  $-1 \div -0.8$ (long-dashed). The photon 
intensities  observed at directions defined by very small 
angles $\theta$ and directions tangent to the massive star limb behaves
differently in different energy ranges. The intensities of GeV photons 
are the highest for directions tangent to the stellar limb and
the lowest for small angles. This is in contrary what is observed 
at TeV energies. Note, that significant amount of secondary photons 
emerges also at the range of angles  $\cos\theta = -1 \div -0.8$. This 
is on the opposite side of the massive star than  the location of the 
compact source of primary photons and/or electrons
(directions obscured by the star!).
Therefore, the soft radiation of a luminous star may work as a kind of 
lens for high energy $\gamma$-ray photons causing the effects of {\it 
focusing of very high energy photons}.

\subsection{Isotropic injection of electrons with the power law 
spectrum}

%___________________________ 
\begin{figure} 
  \vspace{10cm} 
\includegraphics{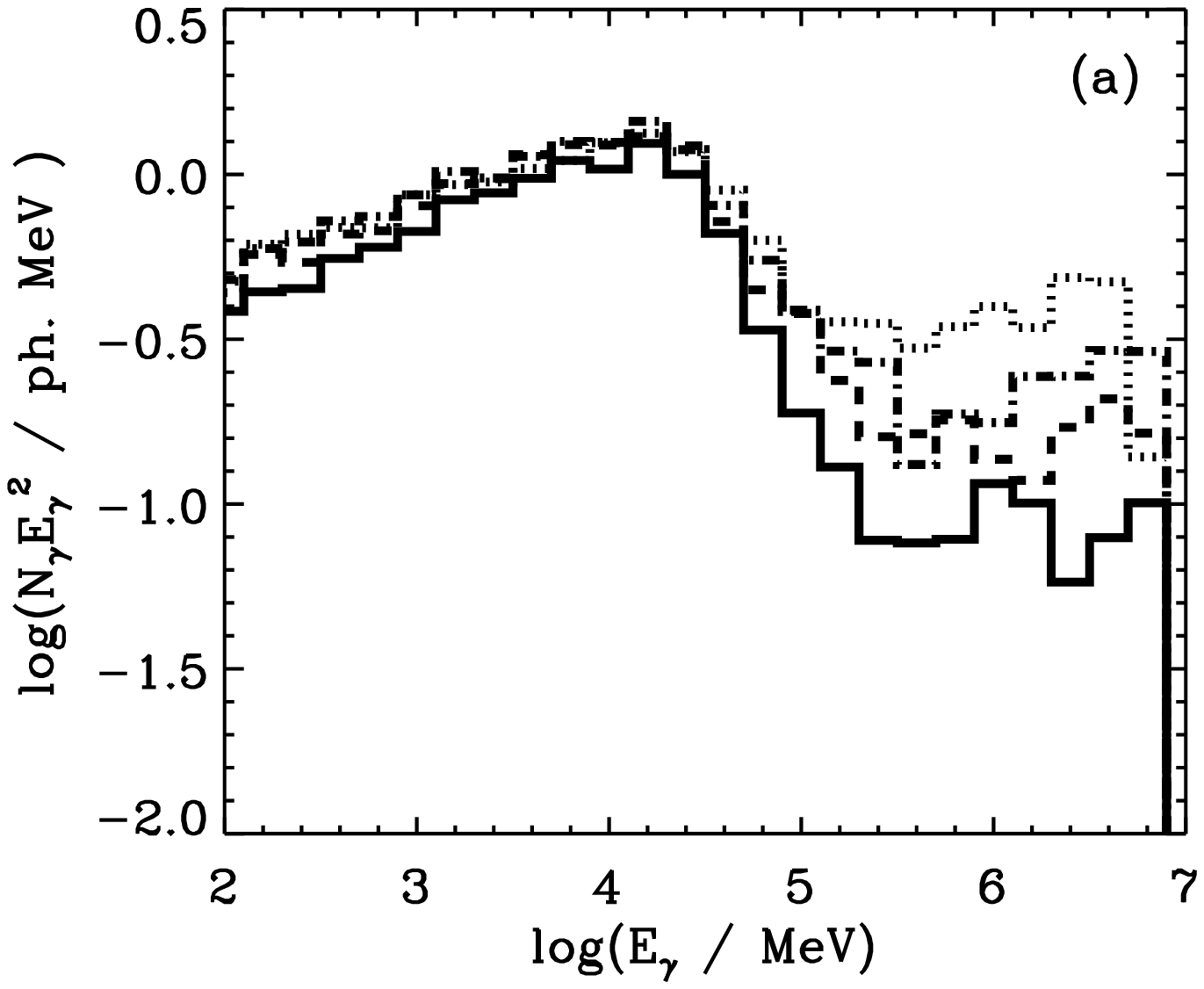}
\includegraphics{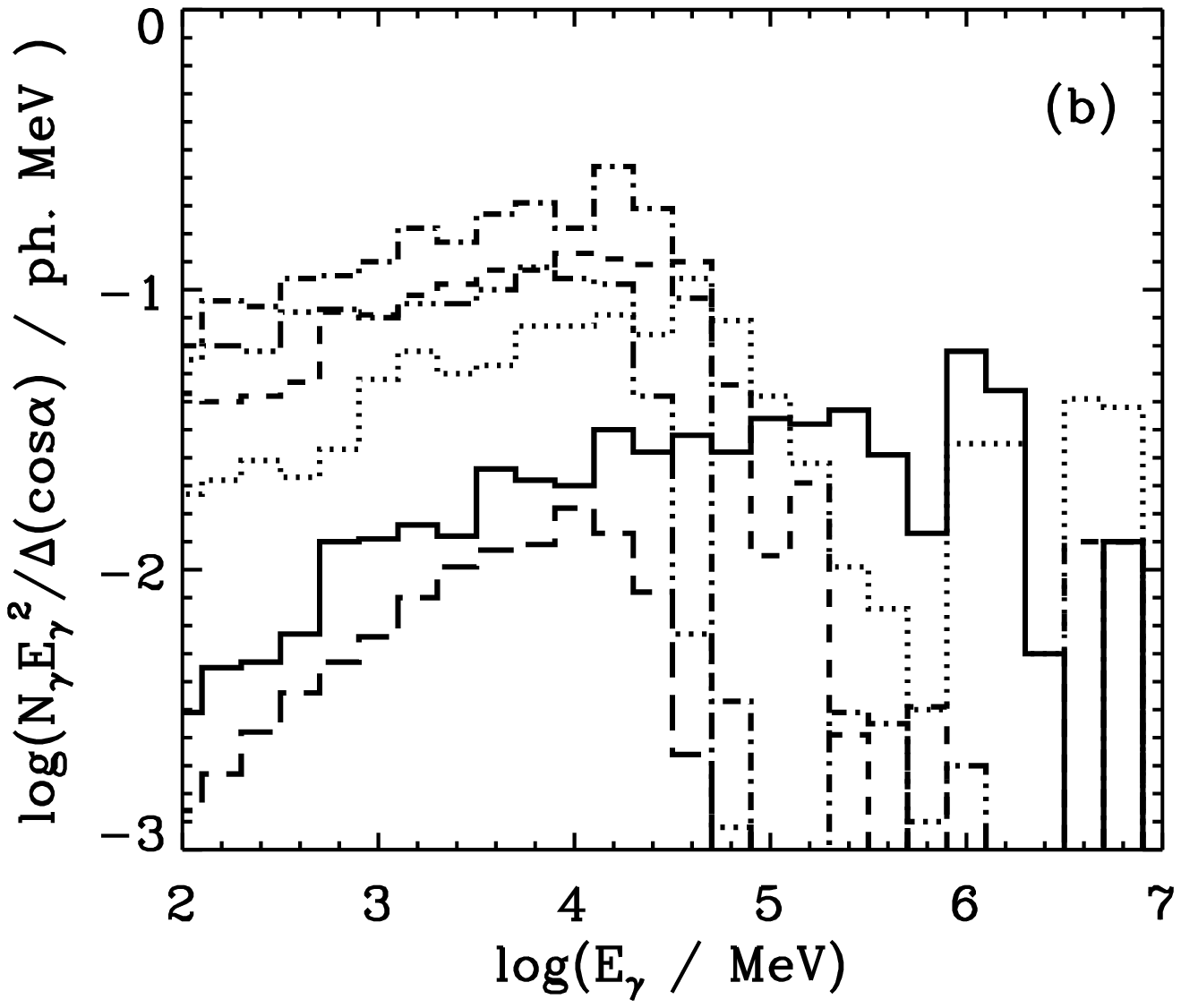}
\caption[]{The $\gamma$-ray spectra observed from the Cen X-3 system in 
the case of isotropic injection of primary
electrons  with the power law spectrum and spectral index -2. 
The spectra integrated over all solid angle are shown in {\bf (a)} for 
different distances of the point source of primary electrons from the 
massive star: $x=1.4r_{\rm s}$ (full histogram), 2 (dashed), 3  
(dot-dashed), and 5$r_{\rm s}$ (dotted). 
The secondary $\gamma$-ray spectra escaping at different range of
angles of observation: cos$\theta = 0.8 \div 1$ (full histogram), 
$0.4 \div 0.6$ (dotted), $0 \div 0.2$ (dashed), $-0.4 \div -0.2$ 
(dot-dashed), $-0.8 \div -0.6$ (dot-dot-dot-dashed), $-1 \div -0.8$
(long-dashed) are shown in {\bf (b)}. } 
\end{figure} 
%___________________________ 

We consider also the case of isotropic injection of electrons with the
power law spectrum. Such electrons can be accelerated at the shock front 
created in collision of the pulsar wind with the surrounding matter 
as proposed in the model by Kennel \& Coroniti~(1984). It is assumed 
that electrons have the power law spectrum with index $-2$ (as expected 
from the theory of acceleration in strong shocks). They are injected 
from the discrete source orbiting the massive star in Cen X-3. 
In the binary system the shock may form relatively close to the 
compact object because the pulsar in Cen X-3 is relatively slow, and the
density of surrounding plasma inside the binary system is high.

The results of calculations of the secondary photon spectra escaping 
from ICS cascade,
after integration over all solid angle, are shown in Fig. 4a
for different distances of the discrete source of primary 
electrons from the massive star. The results show that the integrated 
spectra in the energy region below $\sim 10$ GeV do not depend 
significantly on the location of the injection distance (in the range 
from the surface up to $5r_{\rm s}$). This effect may result from 
our assumption on the local capturing of secondary $e^\pm$ 
pairs by the random
magnetic field. If the dominant magnetic field has ordered structure 
inside the binary then the propagation effects of pairs may be important. Such
more complicated cascade scenario is out of the scope of this paper and
will be discussed in the future work. The photon intensity decreases 
drastically at TeV energies for injection distance closer to the massive 
star surface.

We investigate also the dependence of the shape of the escaping 
spectrum of secondary  $\gamma$-rays on the observation angle $\theta$, 
for
distance of  injection $x = 1.4r_{\rm s}$ (see Fig. 4b). 
The features of these angular
dependent spectra are similar to these ones described above 
for monoenergetic injection of electrons. The highest intensities at TeV 
energies are observed at small angles $\theta$ and the lowest 
intensities at directions behind
the massive star. The highest intensities at GeV energies are for 
directions tangent to the massive star limb and the lowest for small 
angles and at directions obscured by the massive star. 

\subsection{Isotropic injection of photons with the power law 
spectrum}

Finally we discuss the case of isotropic injection of photons with the
power law spectrum and spectral index $-2$. We
show  in Fig. 5 the spectra of escaping $\gamma$-rays for: 
different distances of the injection place from the massive star 
(Fig. 5a);
separately, the escaping spectra of primary $\gamma$-rays and secondary
$\gamma$-rays for the injection place at $x = 1.4r_{\rm s}$ 
(Fig. 5b); and the angular dependence of secondary $\gamma$-ray
spectra on the  observation angle $\theta$ (Fig. 5c). 
General features of these
spectra are very similar to the escaping spectra produced by 
electrons with the power law spectrum. The significant differences 
appear at low energies (below a few GeV) and at high energies (above 
$\sim 1$ TeV), and
results from the contribution to the escaping spectrum  from the primary
$\gamma$-rays which do not cascade in the radiation of the massive star 
in Cen X-3. The escaping primary $\gamma$-rays flattens the spectrum at
energies below  a few $10^4$ Gev (spectral index close to 1.9), and 
dominates the angle integrated spectrum at TeV energies.

%___________________________ 
\begin{figure*} 
  \vspace{5.5cm} 
\includegraphics{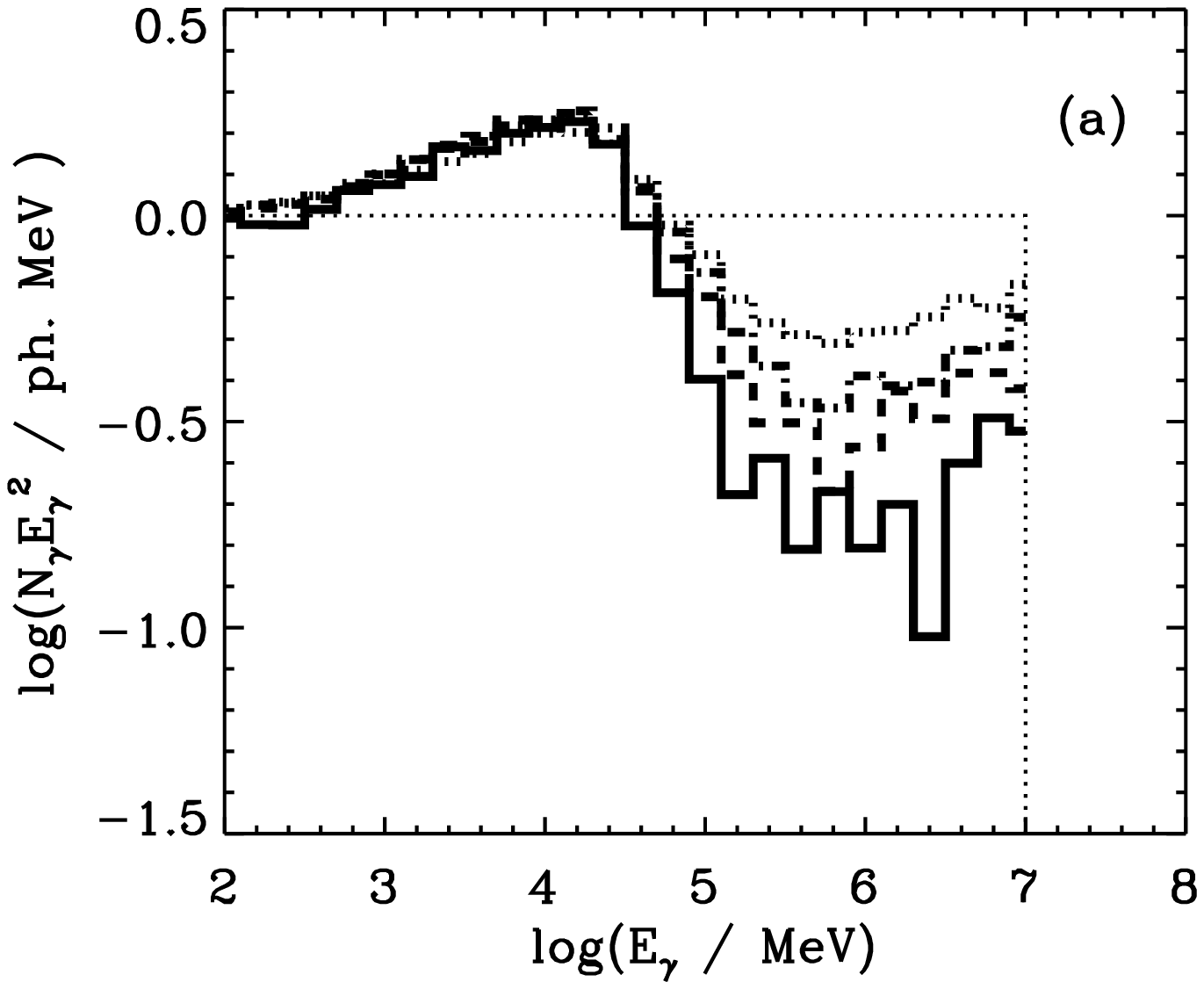}
\includegraphics{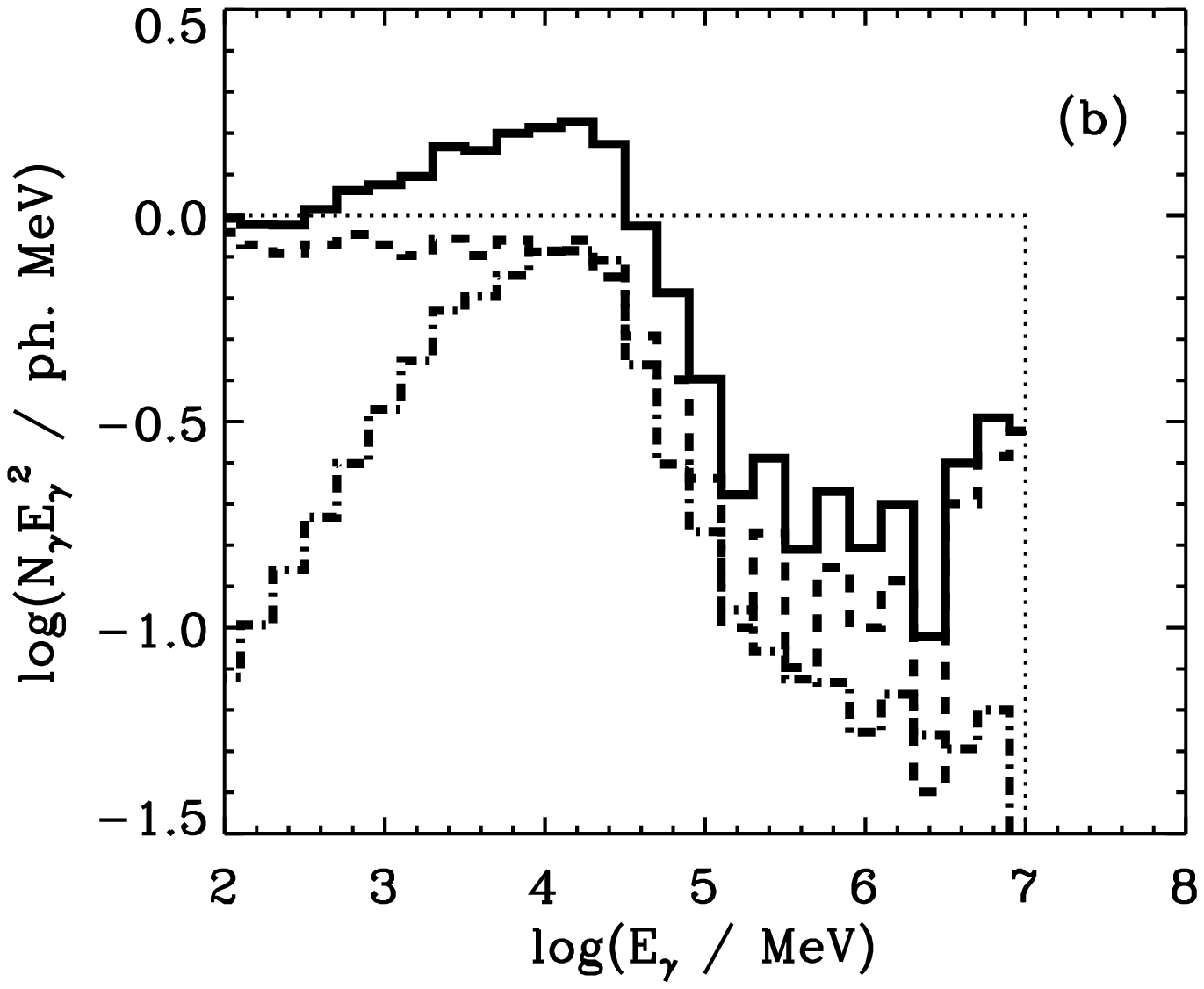}
\includegraphics{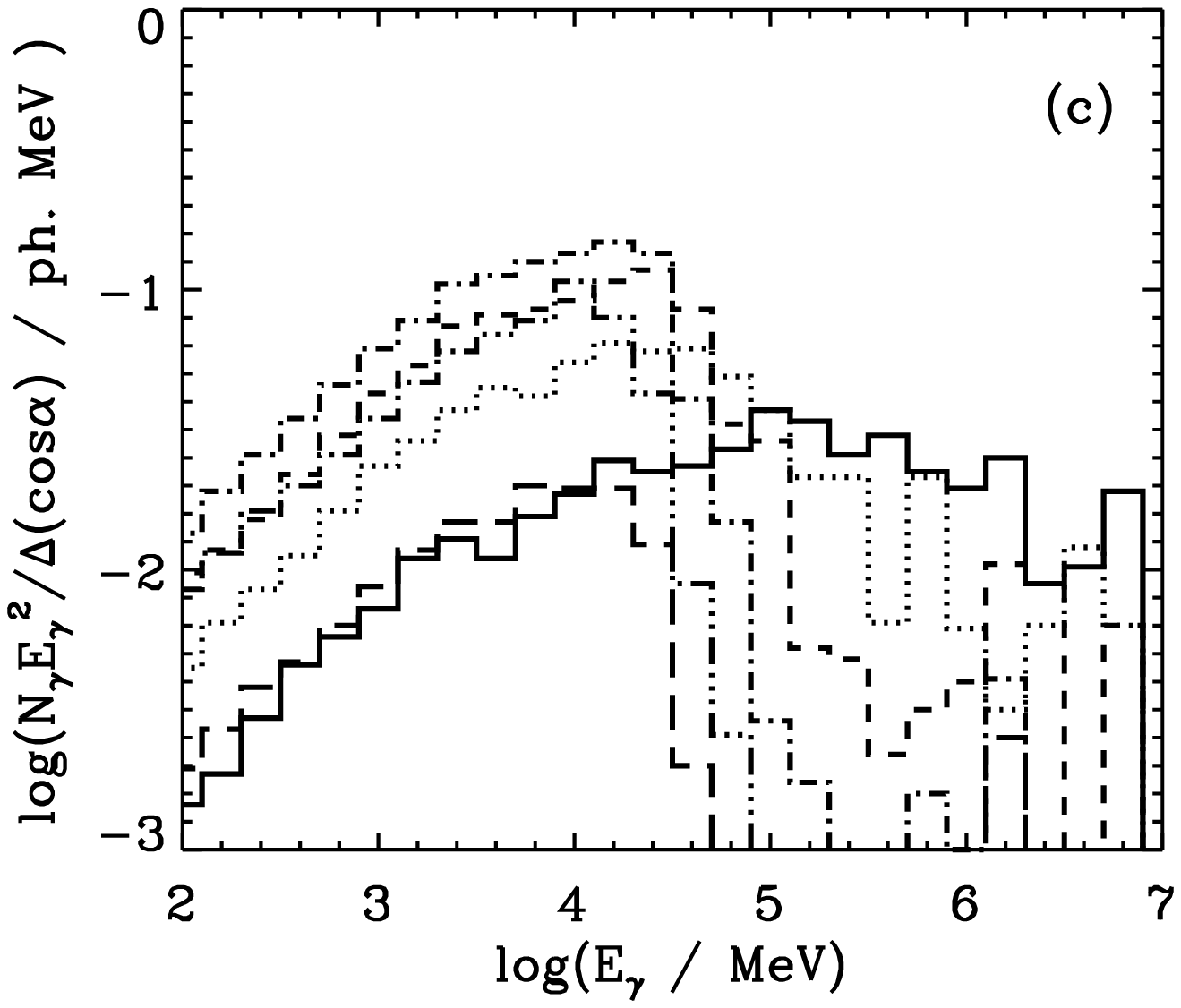}
\caption[]{The spectra of $\gamma$-ray photons escaping 
from Cen X-3 system in the case of isotropic injection of primary 
photons with the power law spectrum and spectral index $-2$, are shown 
in figure {\bf (a)}
for different distances between the center of the massive star and the
injection place: $x=1.4r_{\rm s}$ (full histogram), 2 (dashed), 3 
(dot-dashed), and 5$r_{\rm s}$ (dotted).
The spectrum of photons injected by the compact object is marked by the 
thin dotted line. In figure {\bf (b)} we show the photon spectra 
for the injection distance $x=1.4r_{\rm s}$ which: 
escaped without interaction (dashed curve); are produced in the 
cascade (dot-dashed curve); and all photons escaping from the system 
(full curve). In figure {\bf (c)} the gamma-ray spectra of secondary 
photons, 
in the case of injection at $x = 1.4r_{\rm s}$, are shown for 
different range of angles:  $\cos\theta = 0.8 \div 1$ (full histogram), 
$0.4 \div 0.6$ (dotted),  $0 \div 0.2$ (dashed), $-0.2 \div -0.4$
(dot-dashed), $-0.8 \div -0.6$ (dot-dot-dot-dashed), $-1 \div -0.8$
(long-dashed). }        
\end{figure*} 
%___________________________ 

%
%
\section{Consequences for gamma-ray escape from Cen X-3} 

In the previous section we investigated general features of the
$\gamma$-ray spectra escaping from the radiation field of a massive
star. In order to have results of calculations which can be directly
compared with the observations we compute the $\gamma$-ray light 
curves expected from Cen X-3 system assuming that the compact object
in this binary system injects $\gamma$-ray photons or electrons with
the power law spectrum. The parameters of the Cen X-3 system, used in
computations, are mentioned in the Introduction. Note that the orbit
of the compact object is almost circular, so then the expected 
light curve should be symmetrical. Therefore we compute only the photon
fluxies for the phases from 0 to 0.5, where the phase is measured from 
the side of the observer.

The $\gamma$-ray light curves of photons escaping from the system in 
the case
of isotropic injection of  electrons (or positrons) with the power law 
spectrum and spectral index $-2$ are shown in Figs.~6a,b for 
photons with energies: above 100 MeV, i.e. the EGRET energy range (a), 
and above 300 GeV, i.e. Cherenkov technique energy range (b). 
The results 
are shown for the cut-offs in the spectrum of electrons at $10^7$ MeV 
(full histogram) and at $10^8$ MeV (dashed histogram). Note however, 
that the case with cut-off at $10^8$ MeV we show only for comparison 
since it may not be completely right. Our assumption on the 
isotropization of secondary $e^\pm$ pairs with the Lorentz factors 
$\sim 10^7$ may not be justified in this case(see Eq.~3).  The light 
curves
show that the $\gamma$-ray flux should change  drastically during $\sim 2.09$
day the orbital period of the system by  at least an order of magnitude.
However the $\gamma$-ray  light curves observed at different energy ranges
behaves completely different.  When the photon flux above 100 MeV increases
from the phase equal 0.  up to the eclipse of the  compact object by the
massive star, which occurs for the 
phase $\sim 0.38$, the photon flux above 300 GeV decreases. This  
is the result of propagation of photons in the anisotropic radiation of  
a massive star as discussed  in details in section~3. In Fig.~7, 
we show the spectra of $\gamma$-rays which should  be seen by the 
observer for different phases of the compact object: 0. (full histogram),
0.15 (dotted), 0.35 (dashed), and 0.5 (dot-dashed). The photon spectra 
above 100 MeV have similar shapes but different intensities. The 
spectra above 300 GeV differs significantly not only in the intensities 
but also by the shape.  

%%___________________________ 
\begin{figure} 
  \vspace{9cm} 
\includegraphics{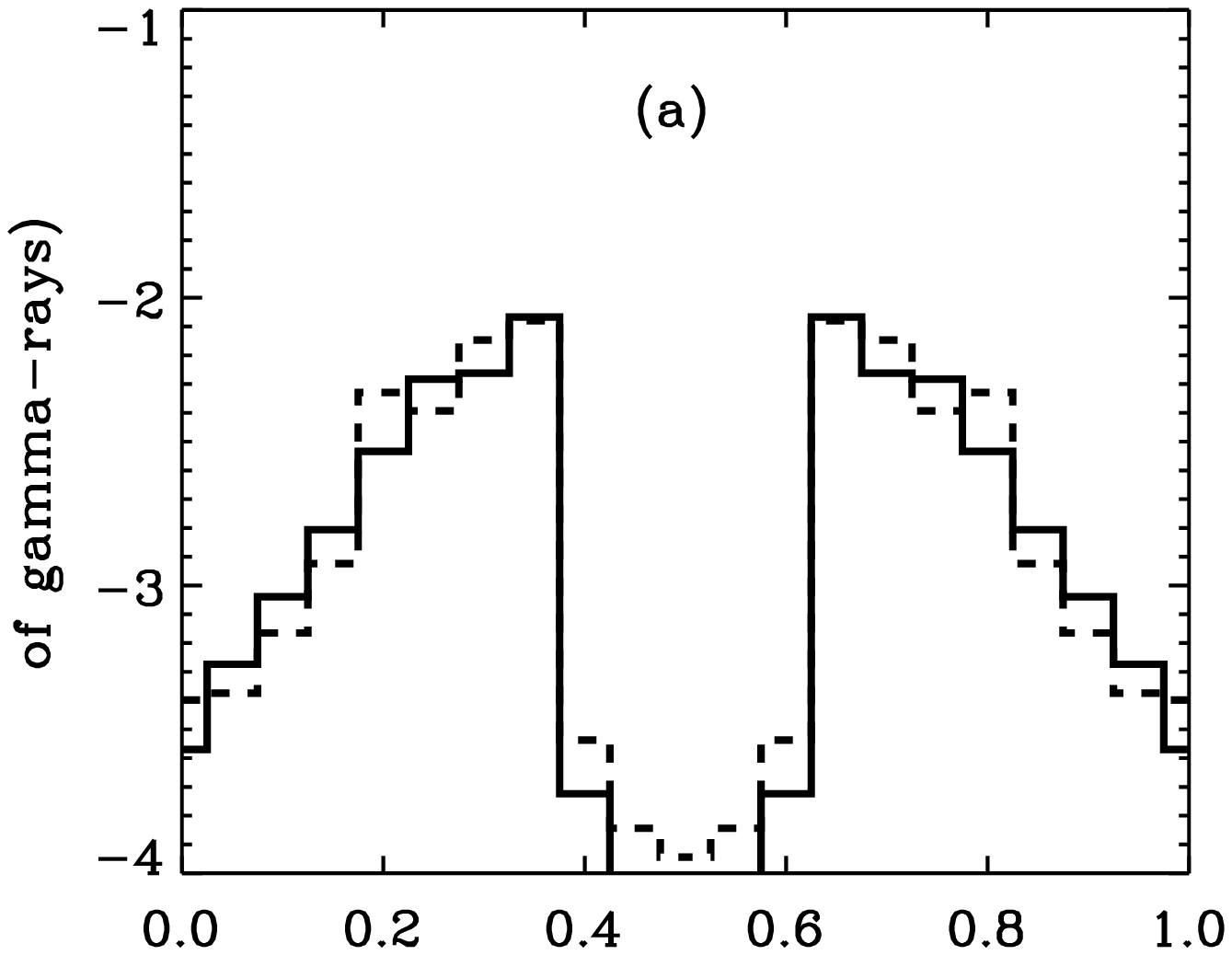}
\includegraphics{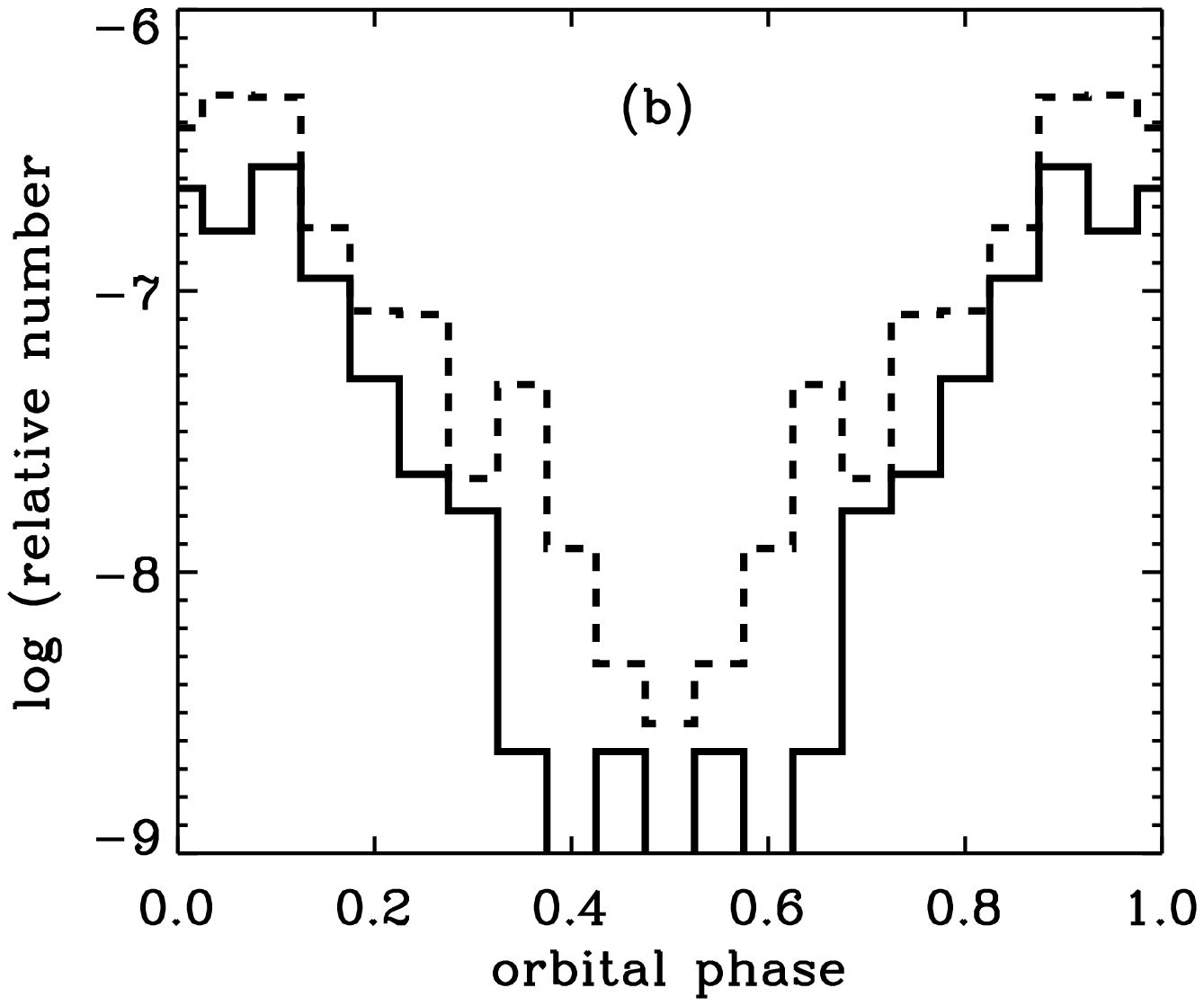}
\caption[]{The light curves observed in $\gamma$-rays at the
inclination angle of the orbital plane of Cen X-3 system equal to 
$70^{\rm o}$ at energies above 100 MeV {\bf (a)}, and 300 GeV  
{\bf (b)} in the case of isotropic injection of electrons with the 
power law spectrum with exponent $-2$. Specific histograms show the 
number of primary photons  which escaped without interaction (dotted),
secondary  photons (dashed)  and all photons (full).}  
\end{figure} 
%%___________________________ 

%
%%___________________________ 
\begin{figure} 
  \vspace{6.cm} 
  \includegraphics{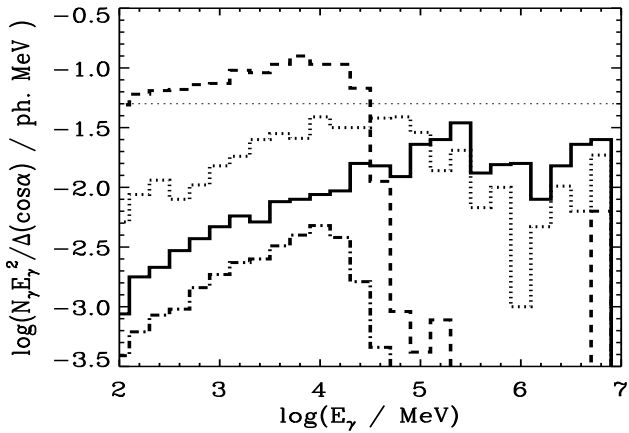}
  \caption[]{Gamma-ray spectra observed from the Cen X-3 system at the
inclination angle $i = 70^o$ for four different phases of the binary
system 0. (full histogram), 0.15 (dotted), 0.35 (dashed), and 0.5
(dot-dashed) within the bin width equal to 0.1. The primary electrons,
with the power law spectrum, exponent $-2$, and the cut-off at
$10^7$ MeV (a) and $10^8$ MeV (b), are injected
isotropically by the compact object.}     
\end{figure} 
%%___________________________ 

We have also computed the $\gamma$-ray light curves in the case of 
isotropic
injection of primary photons with the power law spectrum and index $-2$ 
(see Figs.~8a,b). Specific histograms in these figures show the light
curves  for all escaping $\gamma$-ray photons (full histograms), primary
photons which escape without cascading (dotted), and  secondary photons
produced in cascades (dashed). As expected the light curves for secondary
photons in the case of injection of primary photons and electrons are 
very similar. However the contribution of escaping primary photons 
to the $\gamma$-ray light curves with energies above 100 MeV dominates 
the secondary photons. Altogether, the
$\gamma$-ray light curves at energies above 100 MeV are very flat with 
the strong decrease for phases between $\sim 0.38 \div 0.62$ resulting 
from the eclipse condition. During the eclipse, the
observer may only detect secondary photons produced in cascades (dashed
histogram in Fig.~8b), but on the level of about an order of 
magnitude lower. The $\gamma$-ray light curves above 300 GeV do not 
differ significantly for the case of injection of primary photons or 
electrons
(compare Fig~6b and ~8b). The contribution of primary 
non-cascading photons dominates only for small  
values of the phase (dotted histogram in Fig.~8b).  From these
computations it becomes clear that  investigation of the $\gamma$-ray light
curves at photon energies above 100 MeV (but not  above 300 GeV) should allow
to distinguish what kind of primary  particles is dominantly produced by the
compact object, photons or  electrons (positrons), provided that these
particles are injected isotropically with the power law spectrum.     

%%___________________________ 
\begin{figure} 
  \vspace{9cm} 
\includegraphics{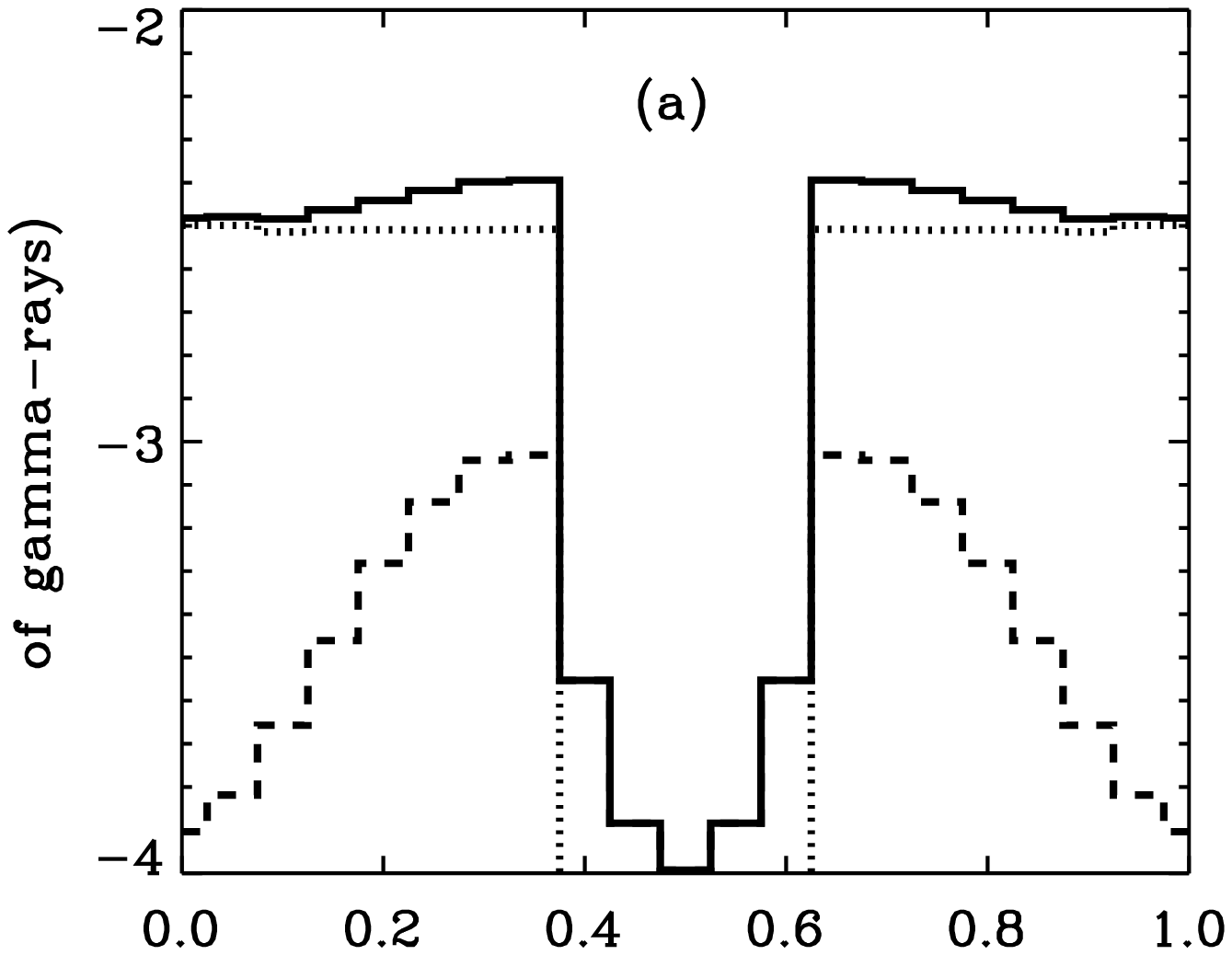}
\includegraphics{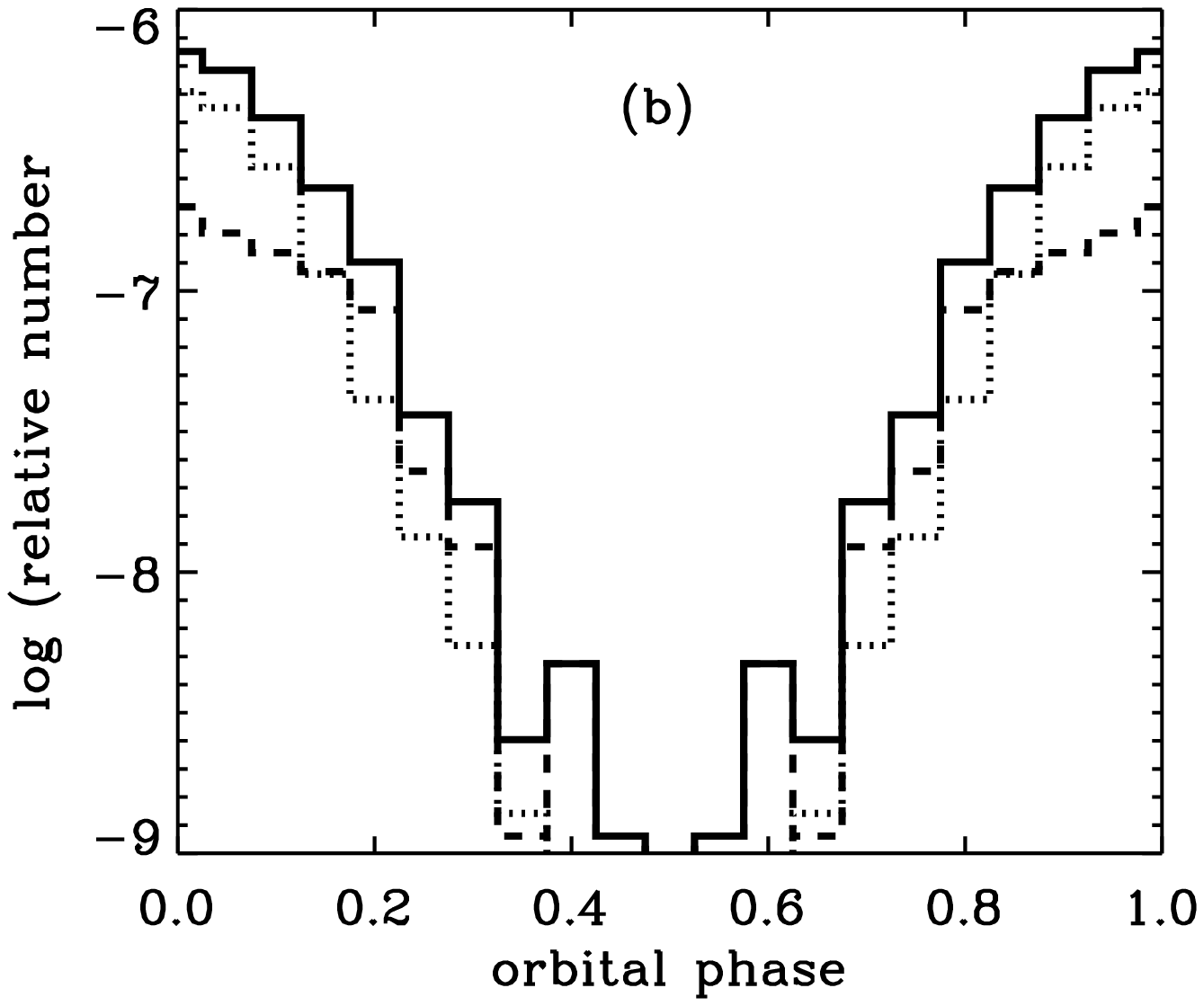}
\caption[]{The $\gamma$-ray photon light curves observed at the
inclination angle $70^{\rm o}$ to the orbital plane in Cen X-3 system 
at two energy ranges: above 100 MeV {\bf (a)}, and above 300 GeV {\bf 
(b)}. The primary $\gamma$-rays, with the power law spectrum 
with exponent $-2$ are injected by the compact object isotropically. 
Specific histograms show: the number of primary photons which escaped 
without interaction (dotted); the secondary photons (dashed); and all 
photons (full).}  
\end{figure} 
%___________________________ 

We also show in Fig. 9a,b the spectra of escaping photons for
different phases of the compact object, separately for secondary cascade
photons and for primary photons which escape without 
interaction.  The photon spectral index below $\sim 
10$ GeV for all escaping photons (primary plus secondary) vary with
phase only in relatively small range, from $-1.8$ to $-2$. The
observed photon fluxies are almost constant. At TeV energies the spectra 
change drastically with the phase of the compact object, similarly to 
the above discussed case of injection of primary electrons. Note, that 
for phase 0.5 (corresponding to the total eclipse of the compact object 
by the massive star) only secondary photons at energies below $\sim 10$ 
GeV can be observed  (Fig. 9a and b).

%
%___________________________ 
\begin{figure} 
  \vspace{12cm} 
\includegraphics{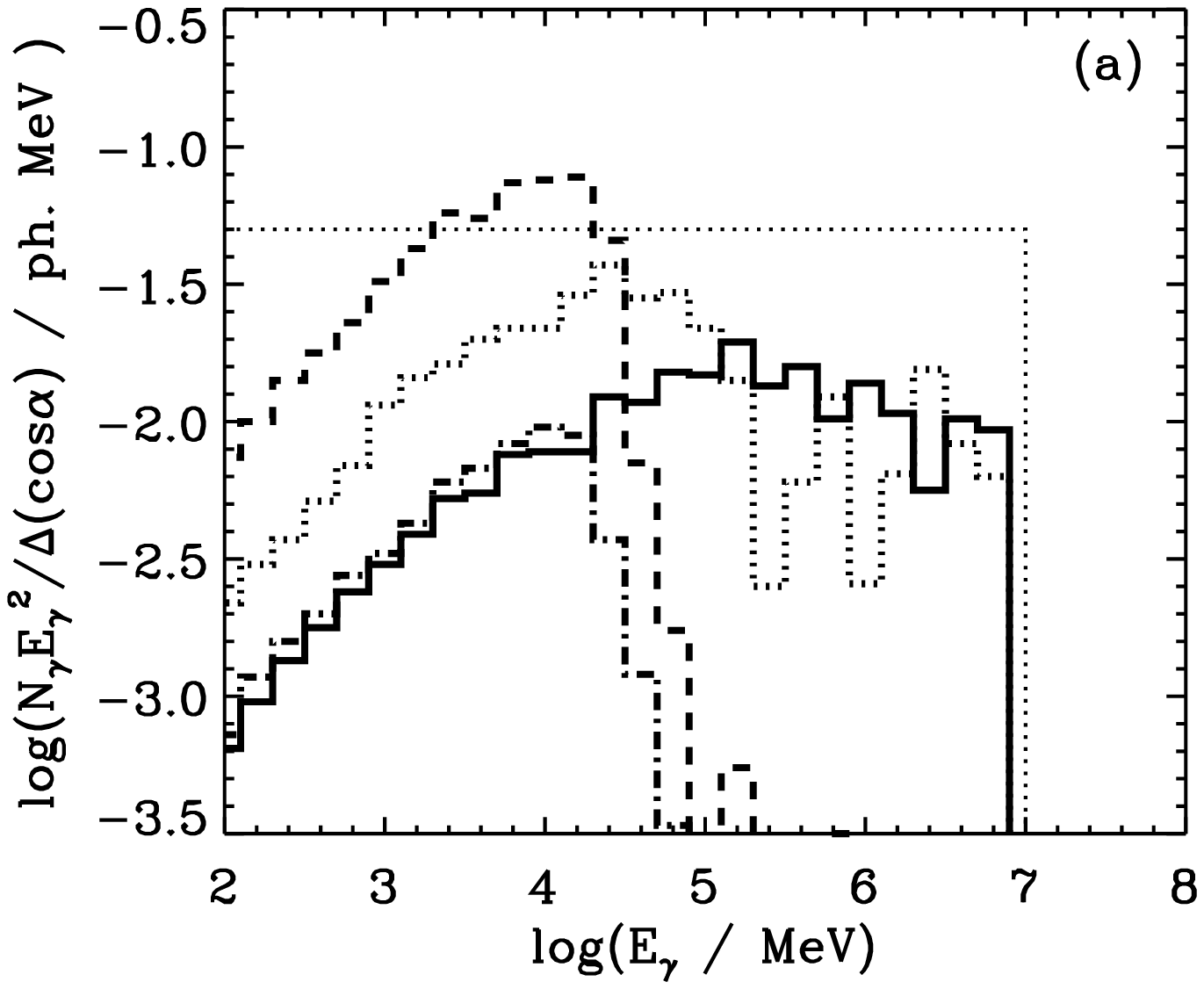}
\includegraphics{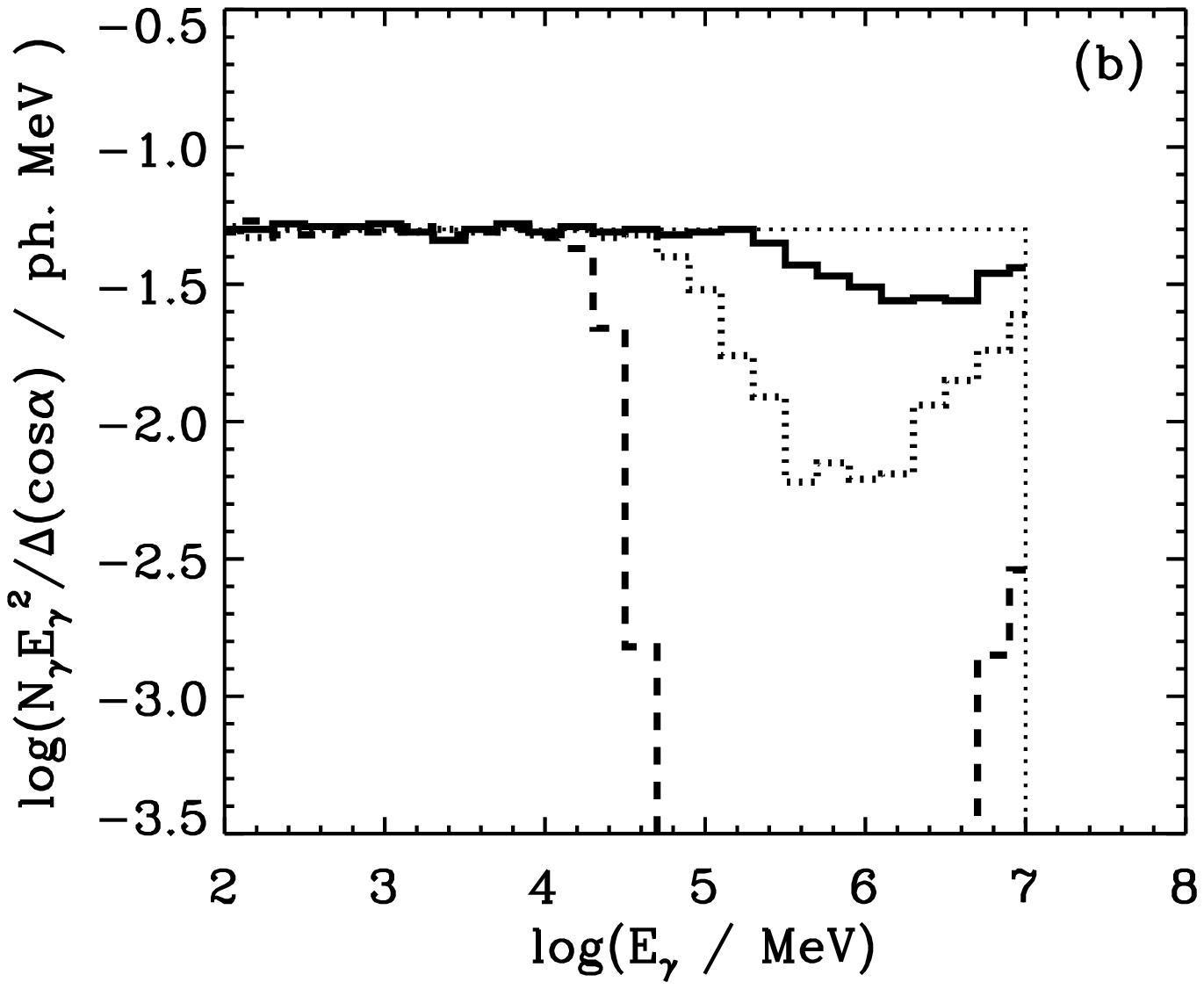}
\caption[]{Gamma-ray spectra of secondary (a) and primary (b) photons 
escaping from the Cen X-3 system at the inclination angle $i = 70^o$ 
for four different locations of the compact source defined by the
phase of the binary system: 0. (full histogram), 0.15 (dotted), 0.35
(dashed), and 0.5 within the bin width equal to 0.1. The primary  
gamma-rays are injected isotropically and have the power law spectrum  
with exponent equal to $-2$.}       
\end{figure} 
%%___________________________ 

As we have discussed in the Introduction, Cen X-3 has been detected in 
GeV and TeV energy range. The emission in GeV energy range can be fitted 
by the power law with the spectral index $-1.81\pm 0.37$  (Vestrand et
al.~1997). This index is consistent with our results for both discussed 
models of isotropic injection of primary photons or electrons with the 
power law spectrum and spectral index $-2$. However, the EGRET 
observations indicate modulation of GeV emission with the pulsar's spin 
period, which should not be observed in the case of injection 
of primary electrons since the escaping
photons at these energies were produced in the cascade process and the
information on the pulsar period should disappear. Therefore the model 
with injection of primary electrons by the compact object seems not work.
The modulation with the pulsar period might be observed in the 
case of injection of primary photons from the compact object, provided 
that the secondary photons do not completely
dominate the primary escaping photons. In fact, this is evident from our 
simulations (see Figs. 9a,b). However, as it is seen in 
Fig.~8a, the photon flux, although constant though most of the
phase range, should drop drastically during the eclipse of the compact 
object by the massive star. This feature has not been observed but also 
can not be rejected by the EGRET observations (Vestrand et al.~1997).

Cen X-3 has been also reported as a source of TeV photons modulated with the
orbital period of the binary system by earlier, less sensitive Cherenkov
observations (Brazier et al.~1990, North et al.~1990). Recent 
observations
report that Cen X-3 is a source of steady emission above $\sim 400$ GeV
(Chadwick et al.~1998). However modulation with the pulsar and orbital 
periods has not been found (Chadwick et al.~1999b). Our calculations 
show that in both models, injection of primary photons and injection of 
primary electrons by the compact object, the modulation of the signal 
with the orbital period should be very clear. In contrary, the 
modulation with the pulsar period should not be observed 
because the secondary cascade photons determines the light curve 
in the case of injection of primary electrons and dominate or give
similar contribution to the light curve in the case of injection of 
primary photons (see dotted histogram in Fig.~8b).

\section{Conclusion}

We considered the cascade initiated by photons or electrons injected 
from the compact object in the radiation field of a massive companion 
in Cen X-3 system, assuming that secondary electrons are isotropised 
by the magnetic field in the binary. It is found that the features of 
the escaping photons from such massive binaries (the light curves, 
photon spectra as a function of the phase of the compact object) may 
allow to distinguish which particles are injected by the compact object, i.e.
photons or electrons. If the cascades are initiated by electrons then 
the escaping secondary cascade photons should not show features of 
modulation with the pulsar period. This seems to be in contradiction 
with the observations of Cen X-3 at energies above 100 MeV. 
Therefore we reject such model. If primary photons are 
injected isotropically with the power law spectrum and spectral 
index $-2$, then the spectrum of escaping photons at
energies below $\sim 10$ GeV is dominated, for most of the phases, by 
the primary non-cascading photons. Then,
the modulation with the pulsar period can be observed. At TeV energies,
the modulation of the photon flux with the 2.09 day binary period should be
strong. The observation of modulation of TeV signal with the orbital 
period of the system have been reported by earlier observations of 
Cen X-3 system (mensioned above)
but not confirmed recently (Chadwick et al.~1999b). We think that this
problem needs further investigation since the sensitivity of present
observations is still rather poor (see the TeV $\gamma$-ray light curve 
in Fig.~3 in Chadwick et al.~1999b).    

If the lack of modulation of the TeV photon flux with the Cen X-3 binary 
period is real, then
a more complicated model has to be investigated. An extended source,
e.g. a shock inside the binary system, injecting primary photons or 
electrons which initiate cascades in the soft radiation of a massive 
star should  be developed. However such computations will require much 
more computing  time in order to get satisfactory statistics, and 
therefore are left for the future work. 

In the present calculations we neglected the X-ray radiation field  
produced by the compact object. Its energy density $L_{\rm X}\approx
10^{38}$ erg s$^{-1}$ (Giacconi et al.~1971) is comparable to the
energy density of thermal photons from the massive star, 
so their photon density is a few orders of magnitude lower inside the 
volume of the binary system. We neglect also the heating effects of the 
massive star by the X-rays coming form the compact object since the 
power emitted by the stellar surface is higher than the X-ray power 
falling on the massive star from the orbital distance of the compact
object equal to 
$1.4$ stellar radii. However note that X-rays, produced in the accretion
disk around a compact object, i.e. close to the production site of 
primary photons, may absorb $\gamma$-rays if they are also produced 
close to the inner disk radius (see e.g. Bednarek~1993). 

%
%
%\acknowledgements{This work is supported by the {\it Polish Committee of
%Scientific Research} (KBN).
% grant  No. 2P03D ?????
%}

%
%

\end{document}